%% file: archiv_final.tex
\documentclass[reprint,natbib209,
 amsmath,amssymb,
 aps,aa
floatfix,
]{revtex4-1}
\usepackage{graphicx}
\usepackage{dcolumn}
\usepackage{bm}
\usepackage{hyperref}
\usepackage{multirow}
\usepackage{wasysym}

\input{config}

\begin{document}


\title{Is the observed 125 GeV Higgs boson expected to be SM-like in the NMSSM?
}

\author{Conny Beskidt}
\email{Conny Beskidt@kit.edu}
\author{Wim de Boer}
\email{Wim.de.Boer@kit.edu}
\affiliation{Dept. of Phys., Karlsruhe Inst. for Technology KIT, Karlsruhe, Germany
}%

\begin{abstract}

In the Next-to Minimal Supersymmetric Standard Model (NMSSM) deviations from the SM signal strengths of the 125 GeV Higgs boson are expected, because of the mixing with the additional singlet-like Higgs boson and/or additional decays into pairs of light particles, like neutralinos, pseudo-scalar Higgs bosons or singlet Higgs bosons.  In this paper the size of the possible deviations and their expected correlations or anti-correlations  between  \textit{bosonic} and \textit{fermionic} final states are analyzed using the efficient parameter scanning technique with complete coverage presented in a companion paper. The regions of parameter space with correlated or anti-correlated deviations of the signal strengths are identified.
  
\end{abstract}

\maketitle

\tableofcontents
\onecolumngrid

\section{Introduction}
\label{Introduction}

After the discovery of the 125 GeV Higgs boson \cite{Aad:2012tfa,Chatrchyan:2012xdj} deviations from SM expectations were found, e.g. in the $\gamma\gamma$ decay, which led to speculations about new physics.\cite{Carena:2012mw,Ellwanger:2011aa,Arvanitaki:2011ck,Gunion:2012zd,Basso:2012tr,Mahmoudi:2012eh,
Baer:2012up,Vasquez:2012hn,Espinosa:2012ir,Choi:2019yrv} However, with higher statistics the measurements became consistent with SM predictions, although the errors are  still significant.\cite{Tanabashi:2018oca} 
In supersymmetric extensions of the Standard Model (SM) one expects one of the light Higgs bosons to have SM-like couplings and branching ratios.\cite{Djouadi:2005gi,Djouadi:2005gj,Martin:1997ns,Carena:2015moc}
In this paper we study in detail what kind of deviations from SM-like signal strengths one can expect in the popular supersymmetric extension of the SM, the Next-to Minimal Supersymmetric SM (NMSSM). A review of the NMSSM can be found in  Ref. \cite{Ellwanger:2009dp}. 

The NMSSM introduces an additional Higgs singlet, which leads to modifications of the Higgs sector compared to the Minimal Supersymmetric SM (MSSM).  The introduction of the singlet has several advantages: the $\mu$ parameter is related to the vacuum expectation values (vev) of the singlet, so it is naturally of the order of the electroweak scale, thus solving the $\mu$ problem, see e.g. Refs. \cite{Kim:1983dt,Miller:2003ay,Ellwanger:2009dp}. In addition, the NMSSM naturally provides a 125 GeV Higgs boson without the need for large loop corrections from multi-TeV stop quarks, since the Higgs mass is increased at tree level by the mixing with the singlet.\cite{Hall:2011aa,Arvanitaki:2011ck,Gunion:2012zd,King:2012is,Kang:2012sy,Cao:2012fz,Ellwanger:2012ke,Beskidt:2013gia} And last, but not least, the superpartner of the singlet provides an electroweak scale dark matter candidate consistent with all experimental data.\cite{Hugonie:2007vd,Kozaczuk:2013spa,Ellwanger:2014dfa,Beskidt:2014oea,Cao:2016nix,
Xiang:2016ndq,Beskidt:2017xsd,Ellwanger:2018zxt} 

The MSSM has five Higgs bosons, while the NMSSM has in total seven Higgs bosons of which two of the scalar bosons and one of the pseudo-scalar bosons are rather light, while the heavier scalar, pseudo-scalar and charged Higgs boson are degenerate in mass if they are well above the gauge boson masses. One of the predicted light Higgs bosons can play the role of the 125 GeV Higgs boson with SM-like couplings.\cite{Carena:2015moc}
However, finding the additional light  Higgs bosons is challenging, since the light scalar and the pseudo-scalar bosons are predominantly singlet-like, so the couplings to SM particles are small.  And finding the additional heavy Higgs bosons is hampered by the fact that the cross sections  decrease fast with increasing mass. 

Instead of  searching for additional Higgs bosons one  can search for hints of the NMSSM by performing precision measurements  of the cross sections and branching ratios of the observed 125 GeV boson, since in the NMSSM deviations from the SM-like properties are expected for several reasons: deviations can be caused by  the extended Higgs sector leading to different Higgs mixing \textit{and/or}   decays of the 125 GeV Higgs boson into  additional Higgs particles of the extended Higgs sector \textit{and/or} decays into invisible particles.

In this paper we study the size of the possible deviations, the correlations between signal strengths of various decay modes and the regions of parameter space, where deviations are expected. We find e.g. regions, where the deviations of decays of the 125 GeV Higgs boson into \textit{bosons} are anti-correlated with the decays into \textit{fermions}, meaning that if one goes up the other ones goes down. But we also find regions with correlated deviations of signal strengths for \textit{bosons} and \textit{fermions}. So measuring signal strengths of decays into \textit{bosons} and \textit{fermions} independently is of great interest for future model building.  For the analysis we use the efficient scanning technique with full coverage, as described in a companion paper, which we call Paper I in the following.\cite{Beskidt:2019mos}

We focus on the semi-constrained NMSSM\cite{Djouadi:2008yj,Ellwanger:2009dp,Kowalska:2012gs},  a well motivated subspace of the general NMSSM allowing to integrate all radiative corrections up to the GUT scale using the renormalization group equations (RGEs). Especially, radiative electroweak symmetry breaking and the important fixed point solutions for the  trilinear couplings are taken into account in this case, thus avoiding trilinear coupling values not allowed  by the solutions of the RGEs, as discussed in Paper I.
As the name fixed point solution indicates, the low energy values are largely  independent of  GUT scale values.
We shortly introduce the NMSSM Higgs sector in Sec. \ref{theory}. The analysis method is described in Sec. \ref{analysis}, including the modifications needed in comparison to the method in Paper I. The results are presented in Secs. \ref{variation} and \ref{single} where we study in detail the possible deviations of the signal strengths compared to the SM expectation.

\section{The Higgs sector in the semi-constrained NMSSM}
\label{theory}

 Within the NMSSM the Higgs fields consist of the two Higgs doublets  ($H_u, H_d$), which appear in the MSSM as well, but together with
an additional complex Higgs singlet $S$.\cite{Ellwanger:2009dp} 
The neutral components from the two Higgs doublets and singlet  mix to form three physical CP-even scalar bosons and two physical CP-odd pseudo-scalar bosons. The mass eigenstates of the neutral Higgs bosons are determined by the diagonalization of the mass matrix, so the scalar Higgs bosons $H_i$, where the index $i$ increases with increasing mass, are mixtures of the CP-even weak eigenstates $H_d, H_u$ and $S$ 
\begin{eqnarray}\label{eq0}
H_i=S_{id}  H_d  + S_{iu}  H_u  + S_{is}  S, 
\end{eqnarray}
where $S_{ij}$ with $i=1,2,3$ and $j=d,u,s$ are the elements of the Higgs mixing matrix.
The Higgs mixing matrix elements enter the Higgs couplings to quarks and leptons of the third generation:
\begin{align}
H_i t_L t_R^c &:  -\frac{h_t}{\sqrt{2}}S_{iu} & h_t &= \frac{m_t}{v \sin\beta},\notag\\
H_i b_L b_R^c &: \frac{h_b}{\sqrt{2}}S_{id} & h_b &= \frac{m_b}{v \cos\beta},\label{coupling}\\
H_i \tau_L \tau_R^c &: \frac{h_\tau}{\sqrt{2}}S_{id} & h_\tau &= \frac{m_\tau}{v \cos\beta},\notag
\end{align}
where $h_t$, $h_b$ and $h_\tau$ are the corresponding Yukawa couplings, $\tan\beta$ corresponds to the ratio of the vev of the Higgs doublets, i.e. $\tan\beta=v_u/v_d$. 
The relations include the quark and lepton masses $m_t$, $m_b$ and $m_\tau$ of the third generation and $v^2=v_u^2+v_d^2$. The couplings to fermions of the first and second generation are analogous to Eq. \ref{coupling} with different quark and lepton masses. The couplings are crucial for the corresponding branching ratios and cross sections for each Higgs boson. 
While the couplings to fermions are proportional to either the mixing element from the up- or down-type state, namely $S_{iu}$ or $S_{id}$ as can be seen from Eq. \ref{coupling}, the couplings to gauge bosons consist of a linear combination of both mixing elements:

\begin{align}
H_i Z_\mu Z_\nu &:  g_{\mu\nu}\frac{g_1^2+g_2^2}{\sqrt{2}} \left ( v_d S_{id} + v_u S_{iu} \right ),\notag\\
H_i W_\mu^+ W_\nu^- &: g_{\mu\nu}\frac{g_2^2}{\sqrt{2}}\left ( v_d S_{id} + v_u S_{iu} \right ),\label{coupling2}
\end{align} 
Thus the couplings to fermions and bosons are  correlated, leading to correlated signal strengths as well. 

\begin{figure}
\begin{center}
\hspace{-1cm}
\includegraphics[width=0.5\textwidth]{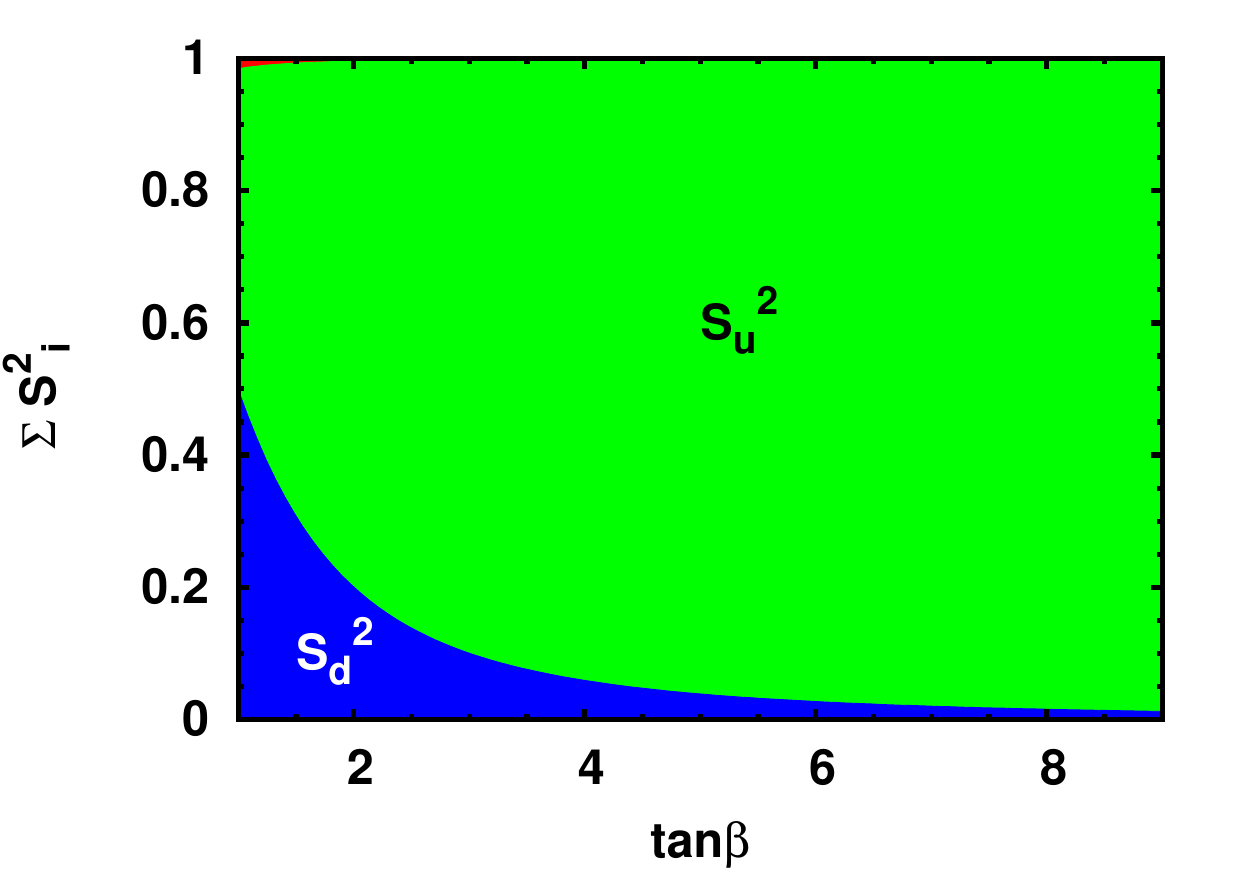}
\caption[]{ 
The sum of the mixing matrix elements squared  for the SM-like Higgs boson as function of $\tan\beta$. The  contributions from $S_d^2$ and $S_u^2$ have been indicated. The singlet component $S_s^2$ is small for most regions of parameter space and is here shown by the thin (red) line at the top of the figure.   
}
\label{fig:1}
\end{center}
\end{figure}

The reduced couplings, meaning couplings relative to the SM couplings, only include  the Higgs mixing matrix elements and $\tan\beta$:

\begin{align}
c_u&=\frac{S_{iu}}{\sin\beta} & c_d&=\frac{S_{id}}{\cos\beta} &c_{W/Z}&= \cos\beta \cdot S_{id} + \sin\beta \cdot S_{iu}, \label{coupling3}
\end{align} 
where $c_{W/Z}$,  $c_u$ and  $c_d$ denote the couplings to vector  bosons, up-type and down-type fermions, respectively.

The loop diagrams  needed for the reduced couplings to gluons $c_{gluon}$ and photons $c_\gamma$ are parametrized as effective couplings  within NMSSMTools.\cite{Das:2011dg}  
If the reduced couplings are SM-like, the Higgs mixing matrix elements  are adjusted such that the reduced couplings are 1 as function of $\tan\beta$, which is possible  if one chooses $S_{iu}\approx \sin\beta$ and $S_{id}\approx \cos\beta$, as is obvious from Eq. \ref{coupling3}.  In this case the couplings to gauge bosons take automatically SM-like values: $c_{W/Z}=\cos\beta\cdot S_{id}+\sin\beta \cdot S_{iu} \approx \cos^2\beta + \sin^2\beta \approx 1$. 
The  components to the sum of the Higgs mixing matrix elements squared of the 125 GeV SM-like Higgs boson  are shown in Fig. \ref{fig:1} as function of $\tan\beta$. 
 One observes that $S_{u}$ is the dominant component  in the linear combination of Eq. \ref{eq0} for the SM-like Higgs boson for $\tan\beta>2$. In contrast, for the heavy Higgs $H_3$ the dominant component is the down-type component, as can be seen from the term $S_{3d}>0.97$ in Table \ref{t3} of Appendix \ref{output}, which will be discussed later in detail. The square of the singlet component   is hardly visible and represented by the thin (red) line at the top of the figure.

The couplings of the 125 GeV Higgs boson can be extracted from  experimentally accessible observables such as  cross sections times branching ratios (BRs).
The cross section times BR relative to the SM values is also known as the signal strength $\mu$ and defined as: 
\begin{align}
\mu_j^i=\frac{\sigma_i \times BR_j}{(\sigma_i \times BR_j)_{SM}}=c_i^2 \cdot \frac{BR_j}{(BR_j)_{SM}}.\label{coupling4} 
\end{align}

So the signal strength is obtained by the  production cross section $\sigma_i$ for  mode $i$ times  the corresponding  BR for  decay $j$, each normalized to  the SM expectation. Normalized cross sections  are called reduced cross sections, which are given by the square of the reduced couplings $c_i$. In the following we focus on the main Higgs boson production modes with the following  reduced couplings $c_i$: the effective reduced gluon coupling $c_{gluon}$ for gluon fusion (ggf), $c_{W/Z}$ for vector boson fusion (VBF) and Higgs Strahlung (VH) and $c_u$ for top fusion (tth). We consider two \textit{fermion} final states (b-quarks and $\tau$-leptons) and two \textit{boson} final states ($W/Z$ and $\gamma\gamma$) for  different production modes.  VBF and VH share the same reduced couplings, so they can be combined to VBF/VH. This leads  to 8 signal strengths in total, namely 4 \textit{fermionic} and 4 \textit{bosonic} signal strengths:   
\begin{align}
\mu_{\tau\tau}^{VBF/VH}&, &\mu_{\tau\tau}^{ggf}&, &\mu_{bb}^{VBF/VH}&, &\mu_{bb}^{tth},\notag\\
\mu_{ZZ/WW}^{VBF/VH}&, &\mu_{ZZ/WW}^{ggf}&, &\mu_{\gamma\gamma}^{VBF/VH}&, &\mu_{\gamma\gamma}^{ggf}.\label{coupling5}
\end{align}

\begin{figure}
\begin{center}
\hspace{-1cm}
\includegraphics[width=0.6\textwidth]{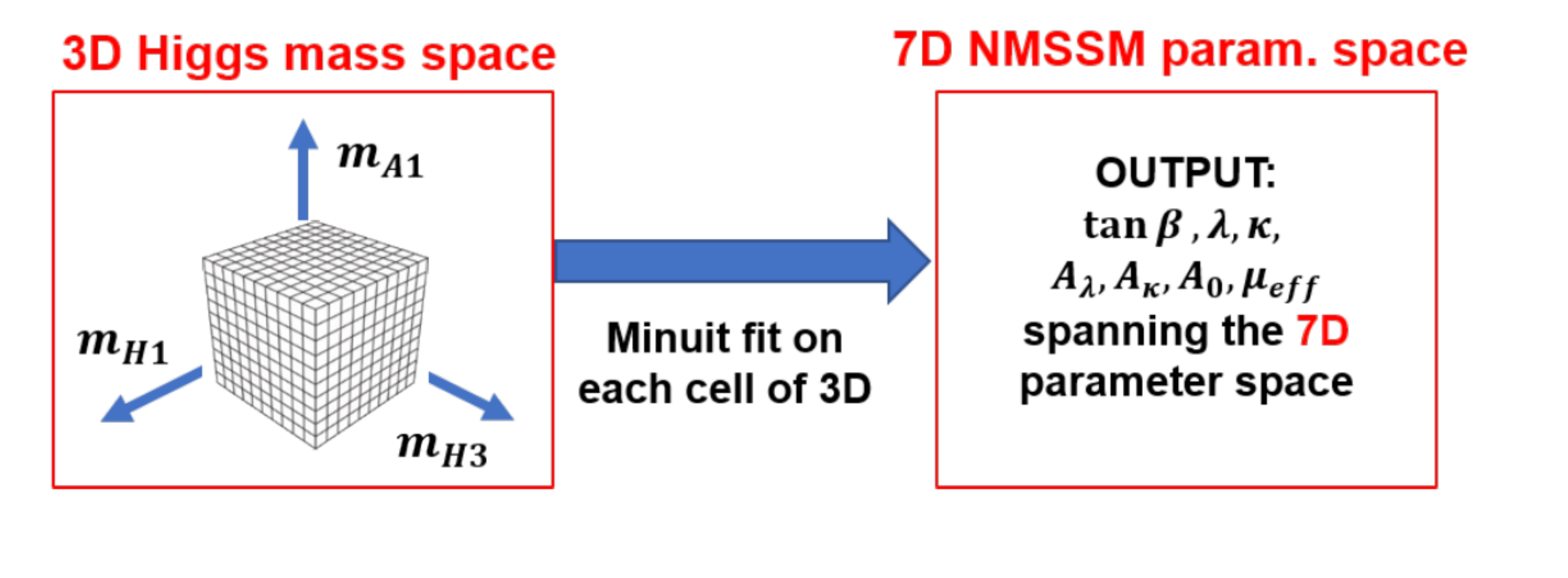}
\caption[]{ Sketch of the sampling technique to determine the allowed NMSSM parameter space. The sampling is done by performing
 a Minuit fit on each cell of the 3D Higgs mass space  (left box) to determine the corresponding 7 free NMSSM parameters (right box). The relation between the NMSSM parameters and the Higgs masses is encoded in NMSSMTools. The second-lightest Higgs boson is chosen to be the 125 GeV Higgs boson, but we repeat the fit in case $m_{H1}$=125 GeV. Then the $m_{H1}$  becomes an $m_{H2}$ axis in the grid on the left. The   3D Higgs mass space can be restricted to the experimentally accessible Higgs masses and the low dimensionality allows a fit to each cell of the grid. This non-random sampling  guarantees   complete coverage of the corresponding 7D NMSSM parameter space. 
}
\label{fig:2}
\end{center}
\end{figure}

\section{Analysis method}
\label{analysis}

The scanning method is described in detail in Paper I.\cite{Beskidt:2019mos} Here we shortly repeat the essential features in order to highlight differences, which are needed  to allow for deviations from SM expectations. In Paper I no deviations from SM-like couplings were assumed in the fit.
The semi-constrained NMSSM\cite{Djouadi:2008yj,Ellwanger:2009dp} has in total nine free parameters: 

\begin{equation}
 \mzero,~ \mhalf,~A_0,~ \tan\beta,  ~ \lambda, ~\kappa,  ~A_\lambda, ~A_\kappa, ~\mu_{eff}.
\label{parameter}
\end{equation}

\begin{table}
\centering
\caption{ \label{t1-diff}
Differences of the fitting procedures from Paper I (middle column) and the one applied in this analysis (last column). 
}
\begin{tabular}{l|p{4.7cm}|p{4.7cm}}
	\hline\noalign{\smallskip}
	Procedure & standard (Paper I)& modified (this paper) \\
	\noalign{\smallskip}\hline\noalign{\smallskip}
	Input & $A_1, H_{1/2}, H_3$, $H_{1/2}=$125 GeV, $\mu_{no-loop}=$1 & $A_1, H_{1/2}, H_3$, $H_{1/2}=$125 GeV, \boldmath$\mu_{sel}=\mu_{theo}$\\
	Output &  $\tan\beta, A_0, A_\kappa, A_\lambda, \mu_{eff}, \lambda, \kappa$ & $\tan\beta, A_0, A_\kappa, A_\lambda, \mu_{eff}, \lambda, \kappa$ \\
	$\chi^2$ constraints & $\chi^2_{H_S}$, $\chi^2_{H_3}$, $\chi^2_{A_1}$, $\chi^2_{H_{SM}}$, $\chi^2_{LEP}$, $\chi^2_{LHC}$ , $\chi^2_{\mu_{no-loop}}$& $\chi^2_{H_S}$, $\chi^2_{H_3}$, $\chi^2_{A_1}$, $\chi^2_{H_{SM}}$, $\chi^2_{LEP}$, $\chi^2_{LHC}$, \boldmath{$\chi^2_{\mu_{sel}}$} \\
	\noalign{\smallskip}\hline
\end{tabular}
\end{table}

The latter six parameters in Eq. \ref{parameter} enter the Higgs mixing matrix at tree level, see Appendix \ref{higgsmixing}, and thus form the 6D parameter space of the NMSSM Higgs sector. 
The coupling $\lambda$ represents the coupling between the Higgs singlet and doublets, while $\kappa$ determines  the self-coupling of the singlet. $A_{\lambda}$ and $A_{\kappa}$ are the corresponding trilinear soft breaking terms. $\mu_{eff}$ represents an effective Higgs mixing parameter, which is related to the vev of the singlet $s$ via the coupling $\lambda$, i.e. $\mu_{eff}=\lambda\cdot s$. 
In addition, we have the GUT scale parameters of the constrained MSSM (CMSSM) $\mzero$ and $\mhalf$ denoting the common mass scales of the spin 0 and spin {1/2} SUSY particles at the GUT scale. The trilinear coupling $A_0$ at the GUT scale is  correlated with $A_{\lambda}$ and $A_{\kappa}$, so fixing it would restrict the range of $A_{\lambda}$ and $A_{\kappa}$ severely. Therefore, $A_0$ is considered a free parameter in the fit, which leads in total to 7 free parameters and thus a 7D NMSSM parameter space. 
For each set of the NMSSM parameters the 6 Higgs boson masses are completely determined. The masses of the heavy Higgs bosons  $A_2$, $H_3$ and $H^\pm$ are approximately equal, so only one of the masses is independent. Furthermore, one of the masses has to be 125 GeV. Then only 3 Higgs masses are free, e.g. $m_{A_1}$, $m_{H_1}$ and $m_{H_3}$. Each set of parameters in the 7D NMSSM parameter space determines a mass combination in the 3D Higgs mass space and vice versa, each mass combination in the  3D mass space corresponds to given regions in the 7D parameter space. Hence it is advantageous to scan the lower dimensional Higgs mass space, especially since this space can be limited to mass ranges accessible to accelerators. Furthermore, the lower dimensionality allows to fit each mass combination in the grid of Fig. \ref{fig:2}. Such a non-random scan guarantees a complete coverage of the corresponding 7D parameter region. A complete coverage is hardly reachable by a random scan in the 7D parameter space because of the high correlations between the parameters, as discussed in Paper I.
 
   The transition of the 3D Higgs mass space to the corresponding 7D NMSSM parameter space can be obtained from a Minuit fit \cite{James:1975dr}, as sketched in Fig. \ref{fig:2} and described in detail in Paper I. The theoretical connection between the NMSSM parameters and masses are obtained from NMSSMTools 5.2.0.\cite{Das:2011dg} We included all available radiative corrections\cite{option}, which is important, since the NMSSM radiative corrections to the Higgs boson can lower the mass by several GeV.\cite{Degrassi:2009yq}
The main contributions and free parameters for the standard fit procedures from  Paper I have been  summarized in Table \ref{t1-diff} in the middle column.

The 4 signal strengths $\mu_{\tau\tau}^{VBF/VH}$, $\mu_{bb}^{VBF/VH}$, $\mu_{ZZ/WW}^{VBF/VH}$ and $\mu_{bb}^{tth}$  do not include loops in which non-SM particles could contribute.  They are generically referred to as $\mu_{no-loop}$. The signal strengths from gluon fusion and/or decay into gammas include loop diagrams at lowest order, so the SUSY particles can contribute leading to deviations from the SM prediction. They are referred to as  $\mu_{loop}$. 
In Paper I  all  $\mu_{no-loop}$ were required to be 1, but
in this analysis we are looking for deviations from  SM expectations, so we do not require $\mu_{no-loop}=1$. Deviations from SM expectations can have  several physical origins, e.g. additional decays of the 125 GeV Higgs boson into lighter particles or modifications of the Higgs couplings via the Higgs mixing elements, which changes the Higgs content. These possibilities will be discussed in detail below.  

Deviations were investigated by a modified fit requiring at least one signal strength to deviate from the SM prediction, i.e. $\mu \neq 1$, which can   be enforced in the fit by replacing the  $\chi^2_{\mu_{no-loop}}$ term in the bottom  line of the middle column of Table \ref{t1-diff} by a term $\chi^2_{\mu_{sel}}=(\mu_{sel} - \mu_{theo})^2/\sigma^2_{\mu}$, which forces  $\mu_{sel}\approx \mu_{theo}$  for a minimal $\chi^2$ value.
For the signal strength deviating from the SM expectation ($\mu_{sel}$) we usually select $\mu_{\tau\tau}^{VBF/VH}$, but other choices from the 8 signal strengths in Eq. \ref{coupling5} can be taken as well, which leads to similar results. 
The fit usually converges to a minimal $\chi^2$ value with    $\mu_{theo}$  selecting the required deviation from the SM expectation. If the fit does not converge or does not reach a small $\chi^2$ value, this means the required deviations cannot be reached in the NMSSM or the selected parameter space does not fulfill all experimental constraints. Such mass combinations are rejected.

In comparison with the standard fit in the middle column of Table \ref{t1-diff} one has 3 constraints less, since in the middle column $\chi^2_{\mu_{no-loop}}$  all four $\mu_{no-loop}$ values were required to be equal 1, while in this analysis the test statistic contribution $\chi^2_{\mu_{sel}}$ requires only $\mu_{sel}=\mu_{theo}$, as shown in the right column of Table \ref{t1-diff}. The reduction of the constraints still leads to converging fits. The reason is the high correlations between the parameters as demonstrated in Paper I. This reduces effectively the number of free parameters. 

The fit is performed for each cell of the grid in Fig. \ref{fig:2}, thus completely scanning the parameter space and checking where  signal strength deviations of the size $\mu_{theo}$ are allowed taking into account the experimental constraints. 
From the fitted parameters in a fit with a single no-loop signal strength required to deviate from 1 the other signal strengths can be calculated and one can see, if they deviate in the same or in opposite directions in comparison to $\mu_{sel}$, which are called correlated or anti-correlated deviations, respectively.

\section{Signal strength deviations of the 125 GeV Higgs boson from SM expectations }
\label{variation}

 The signal strengths $\mu$ are given by the product of cross sections and BRs, see Eq. \ref{coupling4}, so deviation from the SM expectation of $\mu=1$ can be obtained by deviations in cross sections or in branching ratios or in both. This leads to  three different cases  for deviations of the signal strengths of the 125 GeV Higgs boson from  SM expectations:

\begin{itemize} 
\item CASE I: \textit{Additional decays.} The decay of the 125 GeV Higgs boson with SM couplings into particles with a mass $m < 0.5 m_{Higgs}$ leads to modifications of the total width, which changes all BRs in a correlated way. This leads to correlated deviations of the signal strengths.
\item CASE II: \textit{Small Higgs mixing between the 125 GeV  and the singlet  Higgs boson.}  A modification of the Higgs mixing matrix elements leads to anti-correlated deviations because the total width stays almost constant. But the coupling to down-type \textit{fermions} and hence the corresponding BRs decrease, which is largely compensated by an increase of the BRs to \textit{vector bosons}, leading to anti-correlated BRs and thus anti-correlated \textit{fermionic} and \textit{bosonic} signal strengths.
\item CASE III: \textit{Strong Higgs mixing between the 125 GeV  and the singlet-like Higgs boson.}  A strong mixing and thus a large singlet component of the 125 GeV Higgs boson can be reached, if both masses are close to each other. This leads to correlated deviations of the signal strengths, because the singlet component of the 125 GeV Higgs boson does not couple to SM particles, so the reduced couplings decrease in a correlated way. Since a different Higgs mixing does not change the BRs significantly, a correlated decrease of the couplings leads to a correlated decrease of the signal strengths.
\end{itemize}

\begin{figure}[t]
\begin{center}
\begin{minipage}{\textwidth}
\begin{tabular}{ccc}
\textbf{CASE Ia} & \textbf{CASE Ib} & \textbf{CASE Ic} \\
       \includegraphics[width=0.33\textwidth]{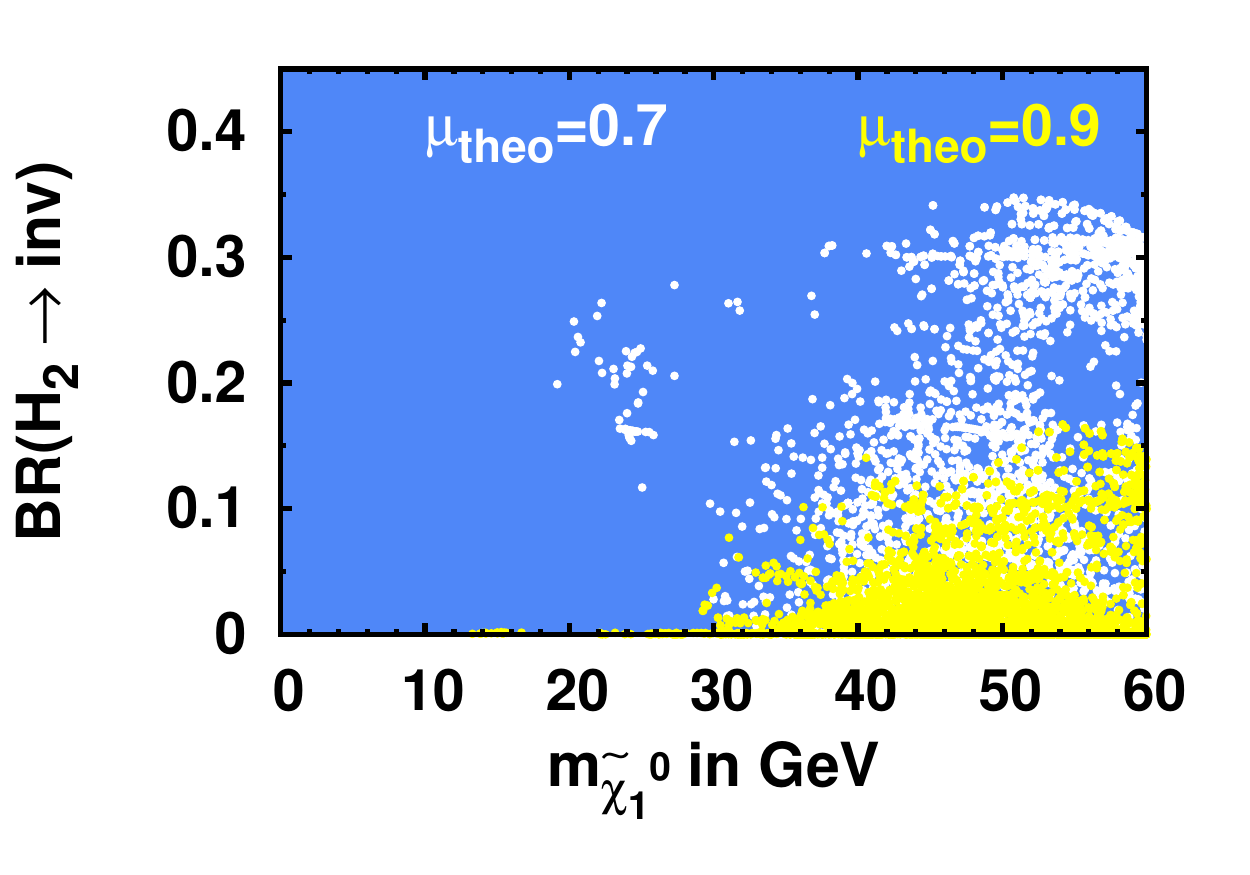} &  
       \includegraphics[width=0.33\textwidth]{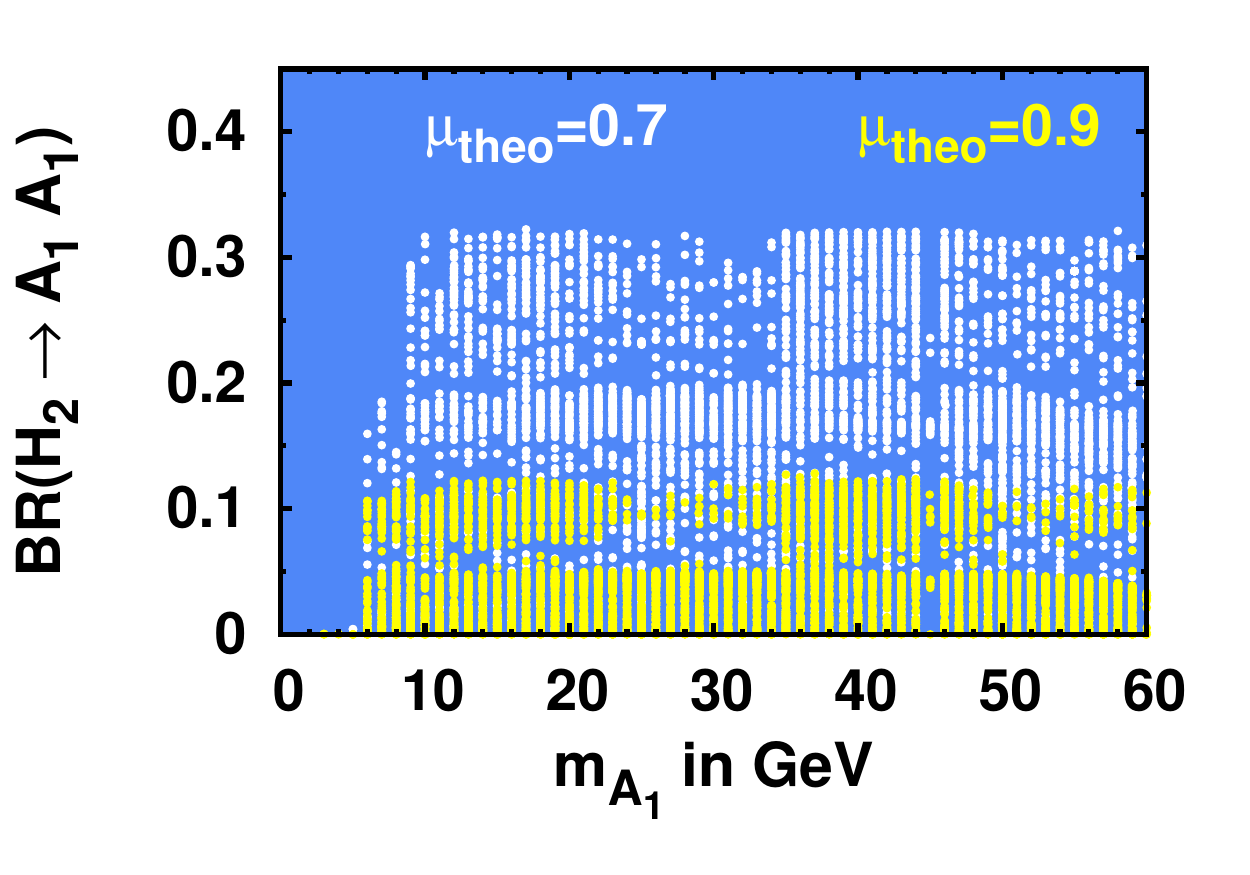} &
       \includegraphics[width=0.33\textwidth]{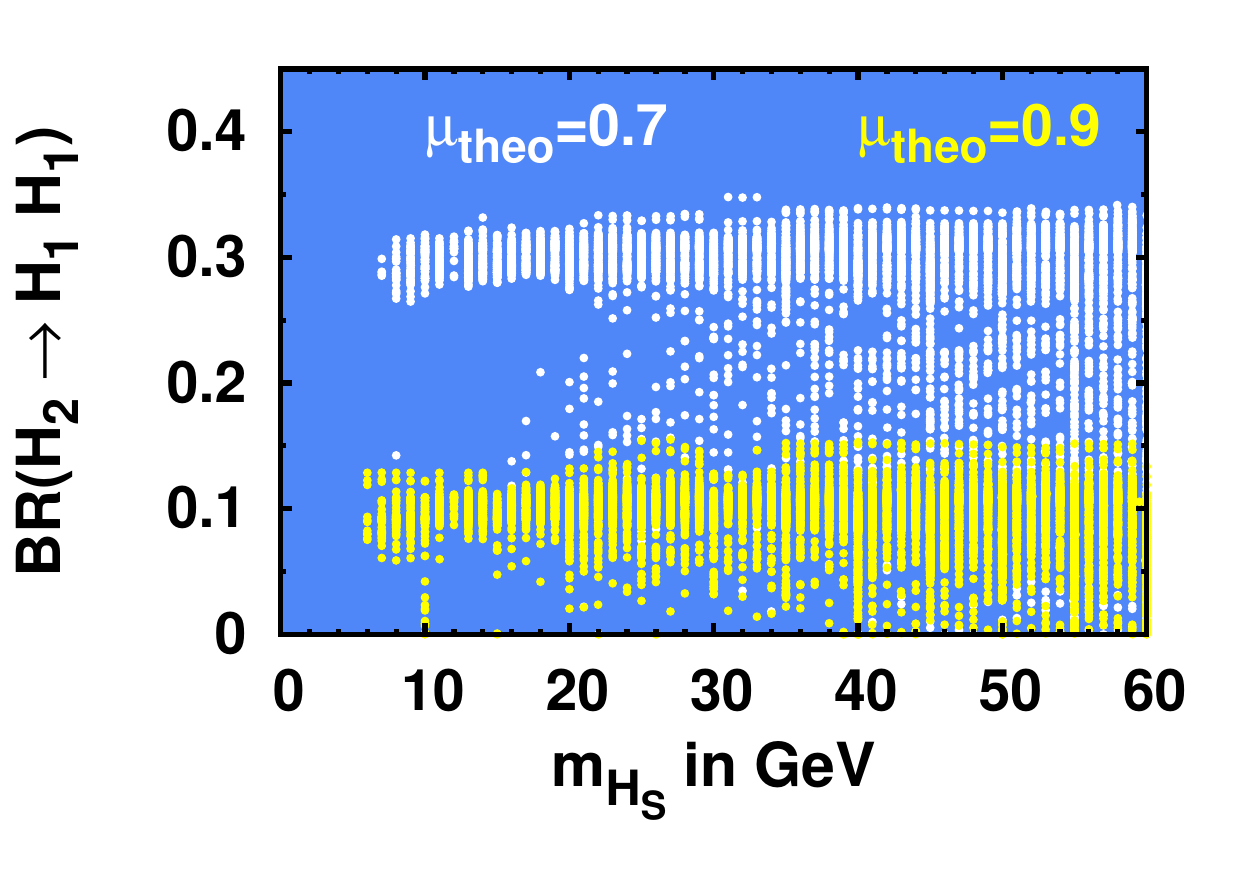} \\
\end{tabular}
\end{minipage}
\caption[]{BRs of the 125 GeV Higgs boson into lighter particles as function of the mass of the lighter particle, which can be neutralinos (CASE Ia), the light pseudo-scalar Higgs boson $A_1$ (CASE Ib) or the lightest Higgs boson $H_1$ (CASE  Ic). The  BRs are shown for $\mu_{sel}=\mu_{\tau\tau}^{VBF/VH}$ and $\mu_{theo}$=0.7 and 0.9.  One observes that the BRs are directly related to the deviations of the signal strengths from the SM expectation, namely the BR is approximately $1-\mu_{theo}$ (see text). 
}
\label{fig:3}
\end{center}
\end{figure}

\begin{figure}[ht!]

\begin{center}
\begin{minipage}{\textwidth}
\begin{tabular}{ccc}
& \footnotesize{\boldmath$\mu_{sel}=\mu_{\tau\tau}^{VBF/VH}\approx\mu_{theo}=0.9$} &\footnotesize{\boldmath$\mu_{sel}=\mu_{\tau\tau}^{VBF/VH}\approx\mu_{theo}=0.7$}  \\
&   \multirow{7}{*}{{\includegraphics[width=0.45\textwidth]{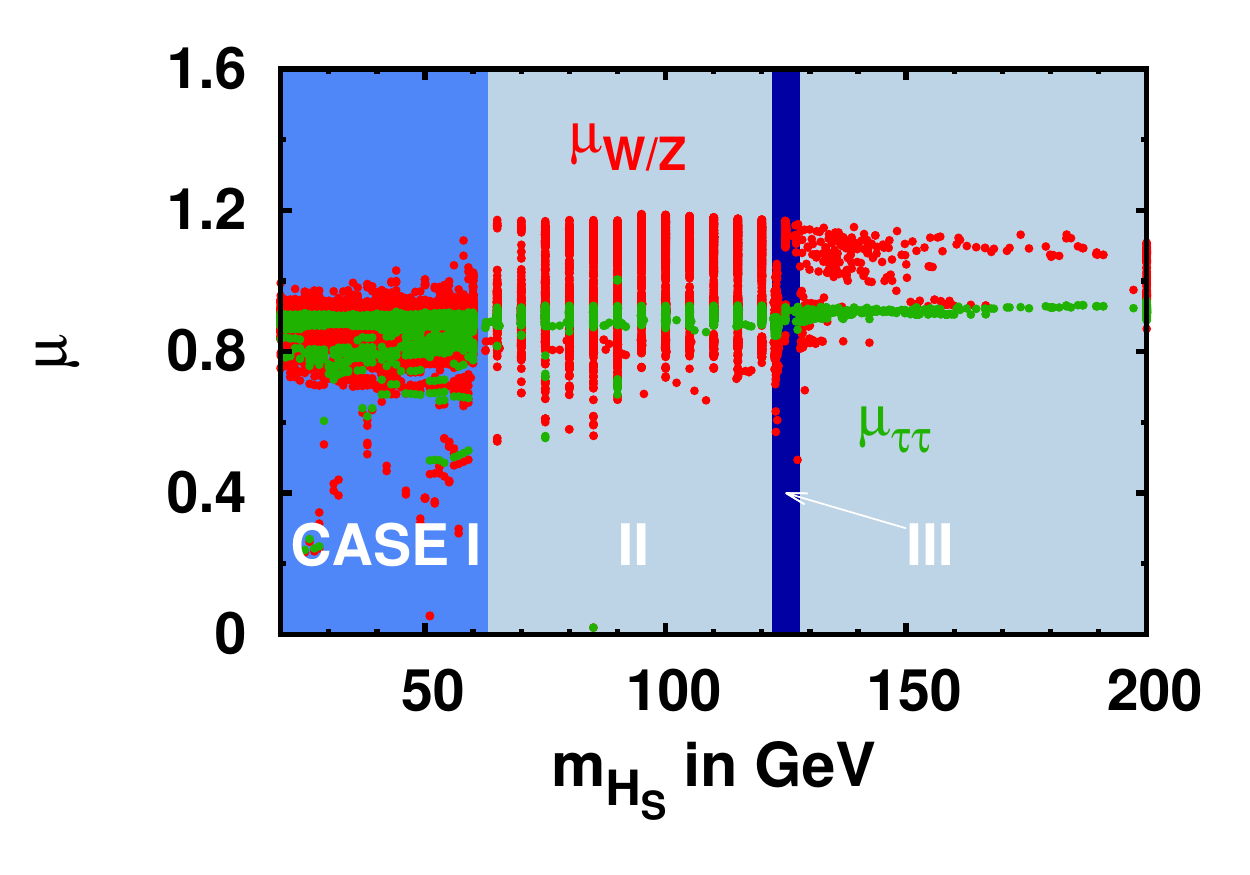}}} & \multirow{7}{*}{{\includegraphics[width=0.45\textwidth]{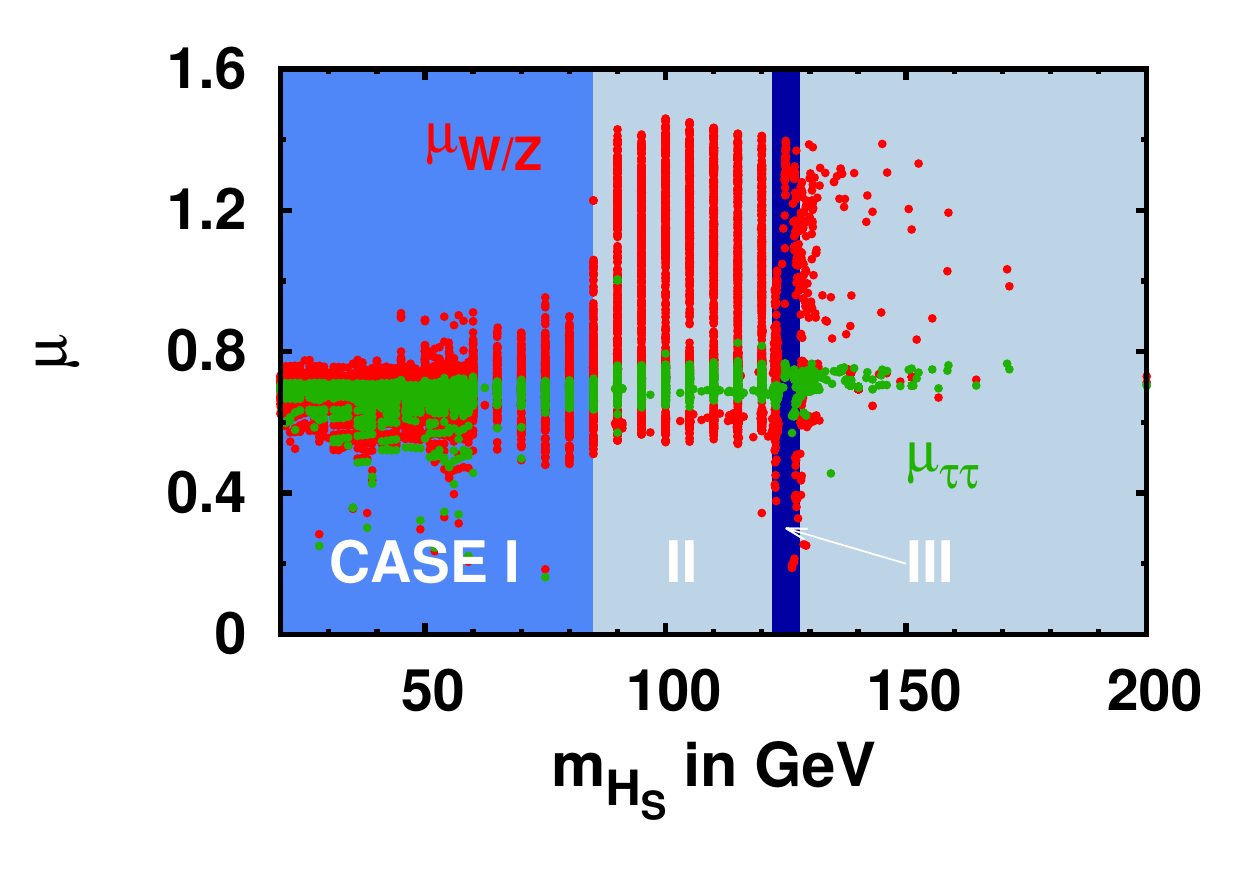}}}  \\
 & &   \\
 & &   \\
 & &   \\
 & &   \\
 & &   \\
 \footnotesize{\textbf{ggf}} & &  \\
 & &   \\
 & &   \\
 & &   \\
 & &   \\
&  &   \\
& &   \\
 & &   \\
&   \multirow{7}{*}{{\includegraphics[width=0.45\textwidth]{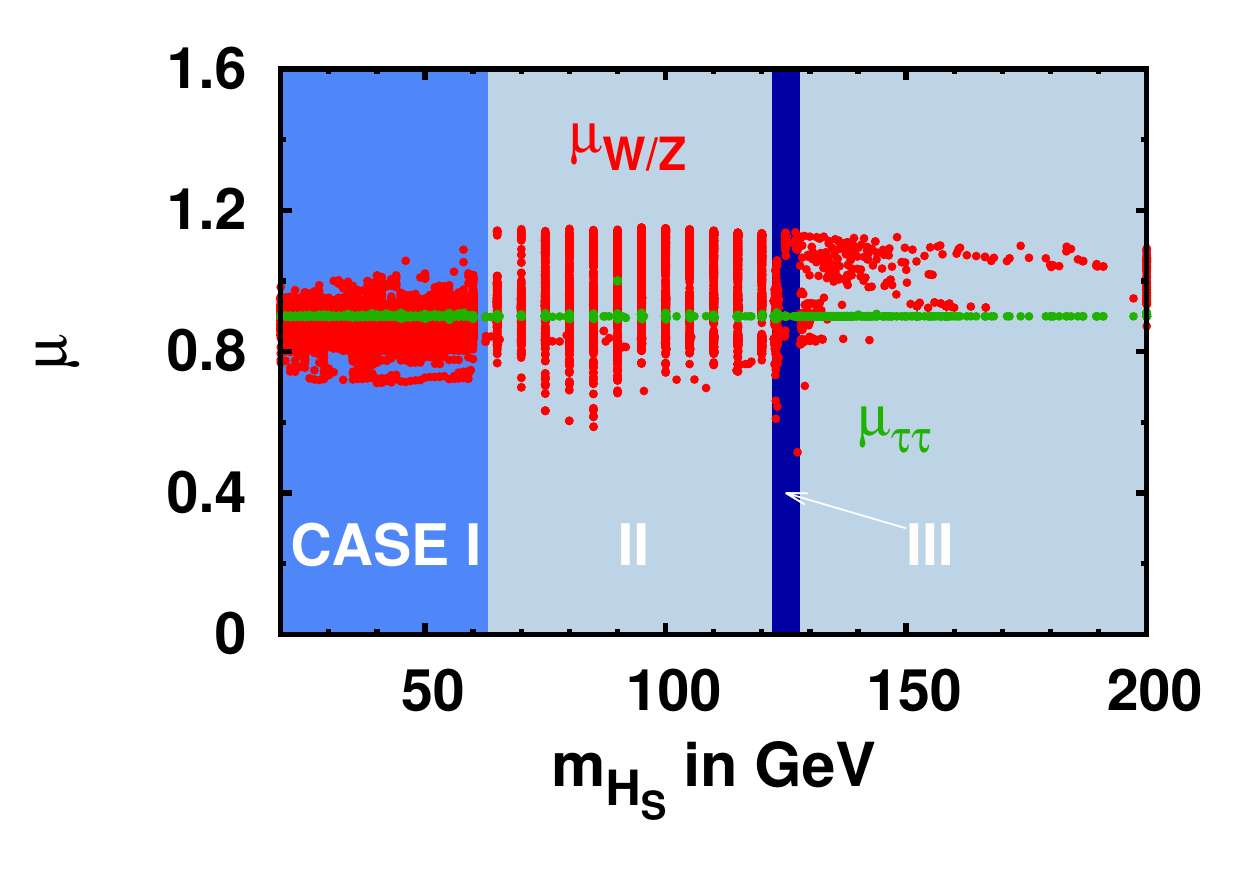}}} & \multirow{7}{*}{{\includegraphics[width=0.45\textwidth]{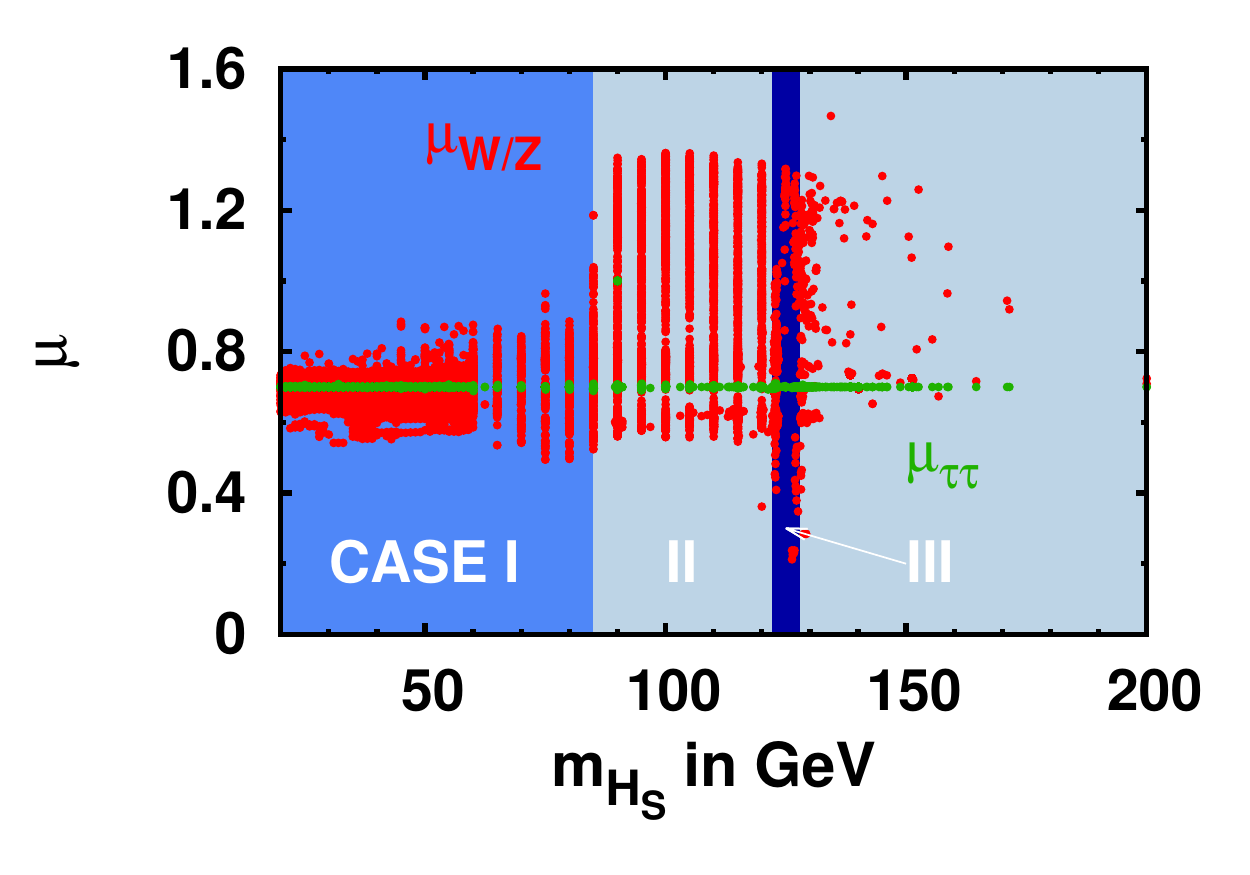}}}  \\
 & &   \\
 & &   \\
 & &   \\
 & &   \\
 & &   \\
 & &   \\
 \footnotesize{\textbf{VBF/VH}} & &  \\
 & &   \\
 & &   \\
 & &   \\
 & &   \\
 & &   \\
 & &   \\
& &   \\
 & &   \\
\end{tabular}
\end{minipage}
\caption[]{Fitted signal strengths $\mu$ as function of the singlet-like Higgs mass for  different values of $\mu_{sel}$, different production cross sections and different final states.
The left(right) panels show results for $\mu_{theo}$ 0.9(0.7), while the top and bottom rows correspond to the ggf and VBF/VH production modes, respectively. The  regions with $m_{H_S}< 60$ GeV  correspond to CASE I and is a combination of CASEs Ia, Ib and Ic in Fig. \ref{fig:3}. The  region with $m_{H_S}> 60$ GeV correspond to CASE II, while the narrow stripe for $m_{H_S}\sim 125$ GeV corresponds to CASE III. One observes correlated signal strengths for \textit{bosons} and \textit{fermions} in CASE I, i.e. they deviate from 1 by the same amount, while for CASE II an anti-correlation of the \textit{bosonic} and \textit{fermionic} signal strengths is observed, i.e. the deviations from 1 go in opposite directions. 
}
\label{fig:4}
\end{center}
\end{figure}

For CASE I deviations are obtained by decays into light neutralinos (CASE Ia), light pseudo-scalar Higgs bosons (CASE Ib) and light singlet-like Higgs bosons (CASE Ic). The BRs are shown in Fig. \ref{fig:3} for each case as function of the mass of the final state particles for two values of $\mu_{theo}$, namely 0.7 and 0.9, respectively. Here $\mu_{sel}=\mu_{\tau\tau}^{VBF/VH}$ was chosen, but similar results are found for other choices of $\mu_{sel}$. The Higgs mass space is scanned on the grid in Fig. \ref{fig:2} in 1 GeV steps. The neutralino mass is calculated  from the fitted NMSSM parameters for each cell in the 3D Higgs mass space.  The fit usually converges with a good $\chi^2$ value meaning that the mass combination  is theoretically allowed and fulfills all experimental constraints.
 Additional decays into light particles increase the total width $\Gamma_{tot}$ of the 125 GeV Higgs boson: $\Gamma_{tot}^*=\Gamma_{tot}+\Gamma_{light} > \Gamma_{tot}$. This leads to a reduction of the BRs of the 125 GeV SM-like Higgs boson, where it is convenient to normalize to the  BR of  the SM. Using the BR into $\tau$ leptons as an example one can write: ${BR_{\tau\tau}^*}/{BR_{\tau\tau,SM}}={\Gamma_{tot}}/{(\Gamma_{tot}+\Gamma_{light})}\approx 1-\Gamma_{light}/\Gamma_{tot}\approx 1-BR_{light}$.  So for CASE I  one basically always finds the deviations from the SM signal strength to be determined by the BR into light particles, i.e. one finds $\mu\approx 1- BR_{light}$.
 This relation approximately holds for the various light particles in the different panels of Fig. \ref{fig:3}. Sometimes it happens that several particles are simultaneously light, since the masses are correlated, which can be seen already from the approximate expressions $m_{A_1} \sim \lambda \mu_{eff}$ and $m_{\tilde{\chi}_0^1} \sim 2 \kappa/\lambda \mu_{eff}$ in Appendix \ref{higgsmixing} or that the mixing between $H_1$ and $H_2$ (CASE II) changes simultaneously with the total width (CASE I). This leads to the broadening of the bands in Fig. \ref{fig:3}, where we summed over all mass combinations of the grid in Fig. \ref{fig:2}.

To study the CASEs II and III, where the deviations of the signal strengths are caused by the Higgs mixing with the singlet, we concentrate on the signal strength as function of the mass of the singlet-like Higgs boson $m_{H_S}$. 
 We select again $\mu_{sel}=\mu_{\tau\tau}^{VBF/VH}$ and $\mu_{theo}=0.9$ or $0.7$. Fig. \ref{fig:4} shows the \textit{fermionic} and \textit{bosonic} signal strengths as function of $m_{H_S}$ for two production channels: ggf (top row in Fig. \ref{fig:4}) and VBF/VH (bottom row in Fig. \ref{fig:4}). 
One observes the following  features:   for $m_{H_S}<60$ GeV in the top left panel $\mu_{W/Z}$ and $\mu_{\tau\tau}$ are both equal to $\mu_{sel}=0.9\approx 1- BR_{light}$, as expected for CASE I. For $m_{H_S}>60$ GeV  the decays into \textit{fermions} and \textit{bosons} are anti-correlated, as expected for CASE II: Since the sum of the partial widths (= total width) stays constant, a change of one partial width has to be compensated by an opposite change in one or more other partial widths (or BRs). Indeed, one observes the signal strengths for $\mu_{W/Z}$ to be above the ones for $\mu_{\tau\tau}$. The effect becomes more pronounced, if one requires larger deviations, e.g. $\mu_{theo}=0.7$, which is shown on the right side of Fig. \ref{fig:4}. Note that  the spread is large  because  the results are shown for all mass combinations in the grid of the Higgs mass space and for different stop masses (determined by different values of $\mzero,\mhalf$).  All cases will be discussed in more detail in the next section, where we do not average over all mass combinations, but consider for each case a representative mass combination, which better shows the salient features. 

\begin{figure}[]
\begin{center}
\begin{minipage}{\textwidth}
\begin{tabular}{cc}
\includegraphics[width=0.45\textwidth]{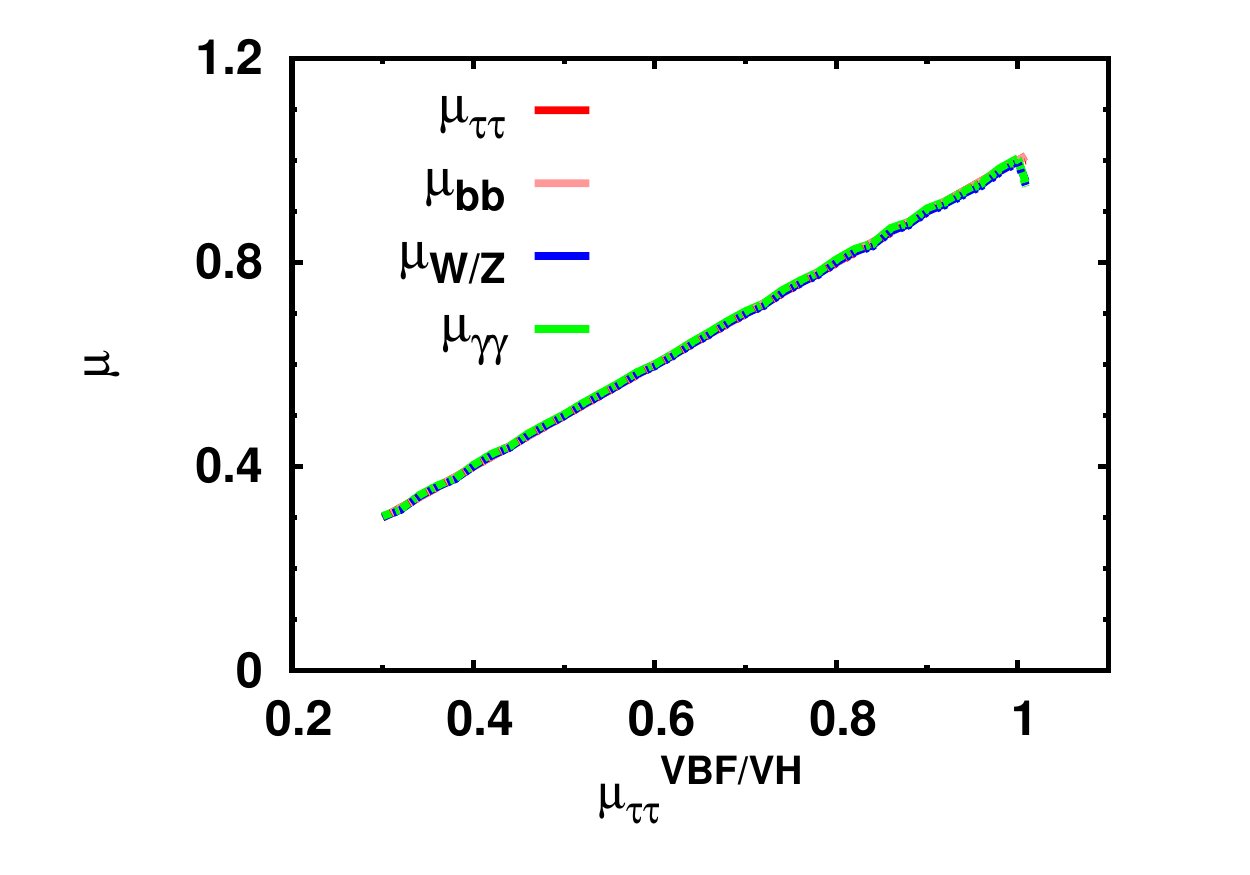} & 
 \includegraphics[width=0.45\textwidth]{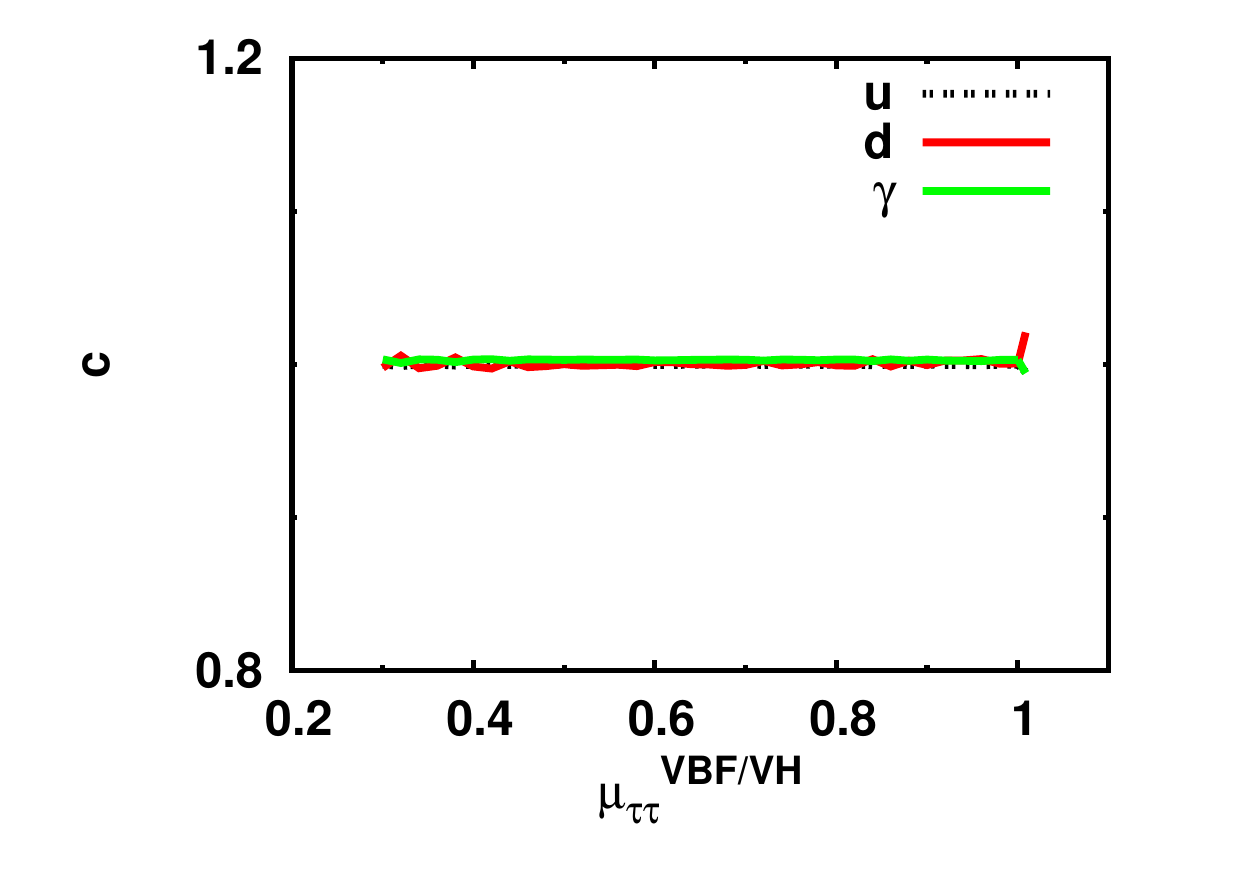} \\
(a) & (b) \\
\includegraphics[width=0.45\textwidth]{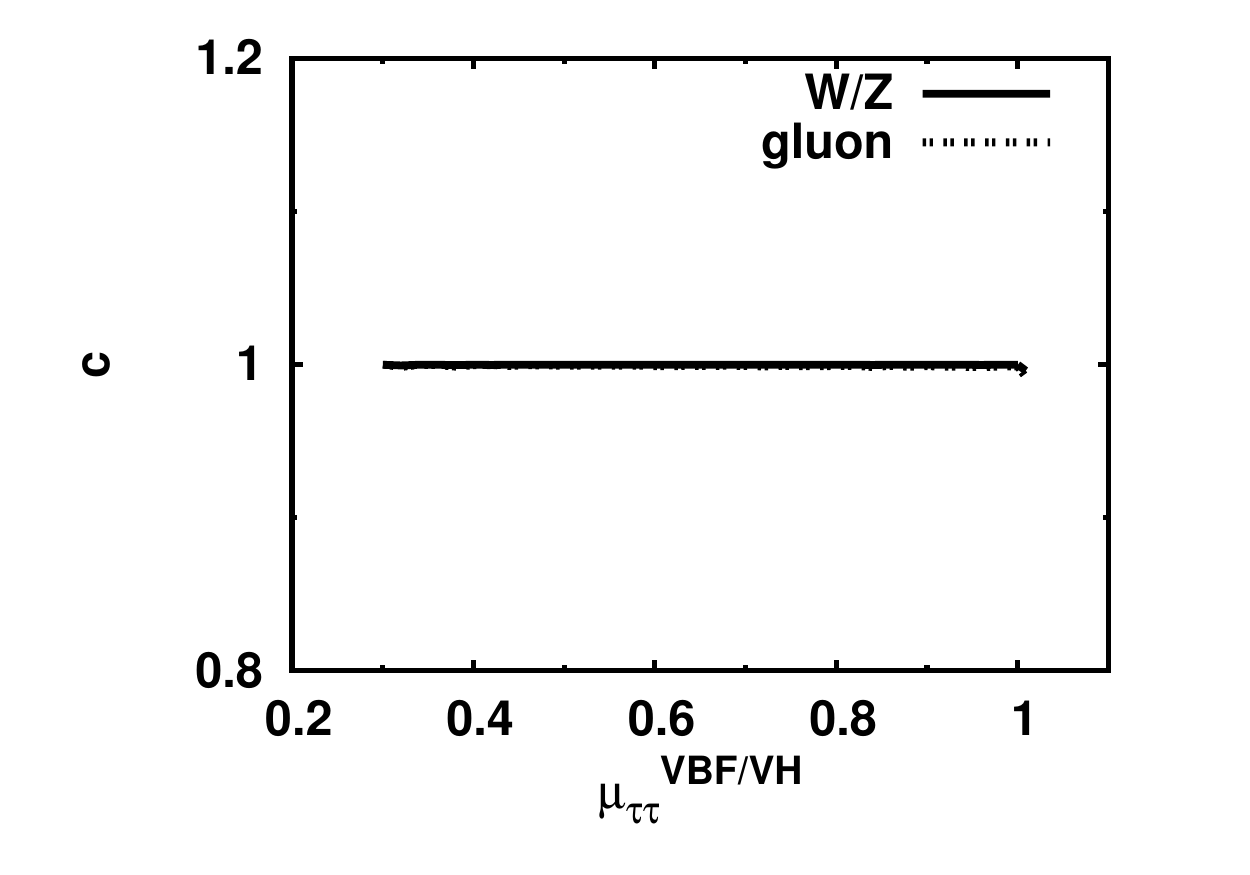} & 
 \includegraphics[width=0.45\textwidth]{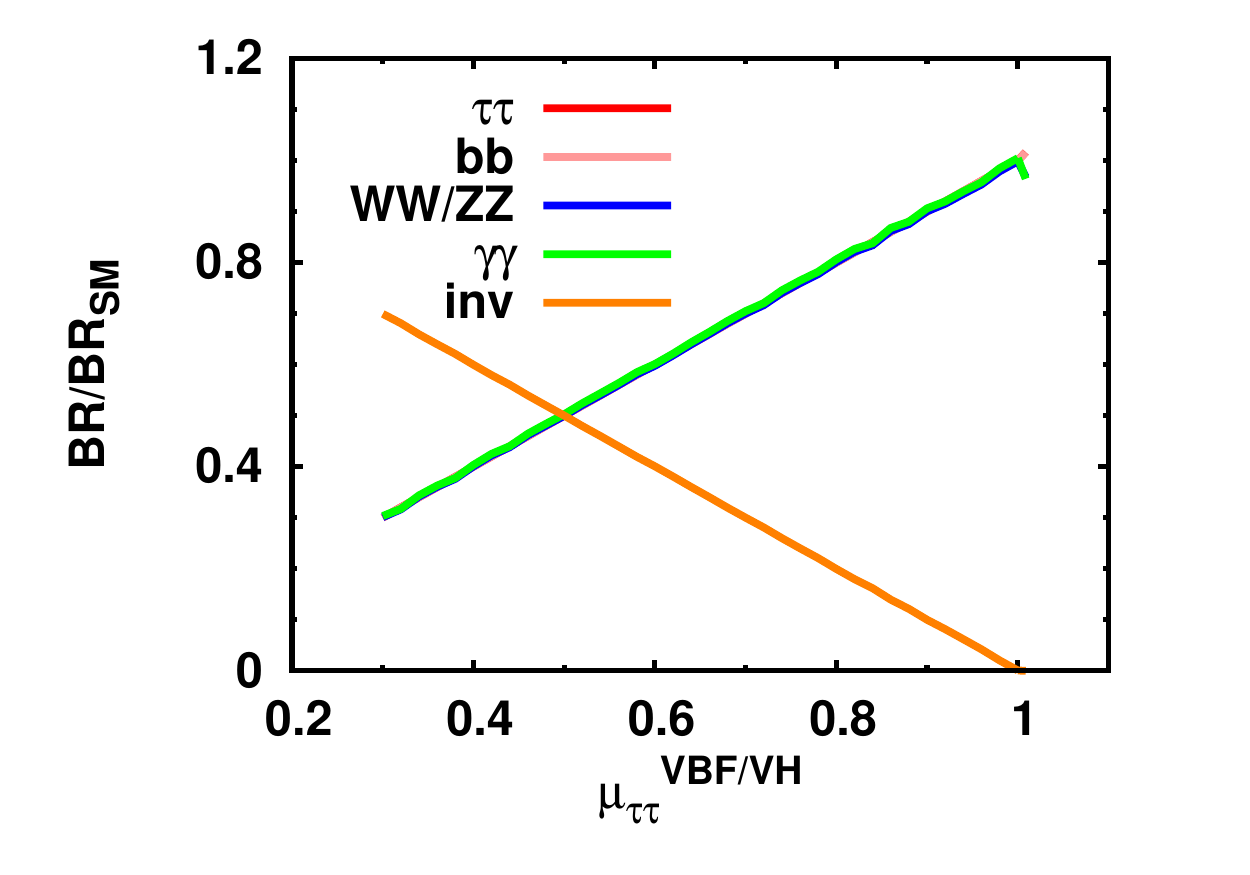} \\
(c) & (d) \\
\end{tabular}
\end{minipage}
\caption[]{The figure  demonstrates  in the top left panel (a) the correlated deviation of all signal strengths from the SM expectation 1 as function of  $\mu_{\tau\tau}^{VBF/VH}$ by the overlapping lines. $\mu_{\tau\tau}^{VBF/VH}$ was required to deviate from the SM expectation by  fitting it to a value $\mu_{theo}<1$ for a specific mass combination characterized by allowed decays into neutralinos (CASE I). The deviations can either come from deviations of the couplings or from the BRs. Here it is clearly caused by the BRs, since all reduced couplings correspond to the SM expectation of 1, as demonstrated in panels (b) and (c).  The allowed neutralino decays increase the invisible width (line with negative slope in panel (d)), which increase the total width and hence reduces the BRs for all other channels in the same way (overlapping lines with positive slope in panel (d)). This leads to the same (=correlated) change in all signal strengths (overlapping lines with positive slope in panel (a)) given the constant couplings in panels (b) and (c), which
  is true for all production modes, as demonstrated by the overlap of the solid and dashed lines in (a) representing the signal strengths for the VBF/VH and ggf production mode, respectively. 
}
\label{fig:5}
\end{center}
\end{figure} 

\begin{figure}[]
\begin{center}
\begin{minipage}{\textwidth}
\begin{tabular}{cc}
\includegraphics[width=0.45\textwidth]{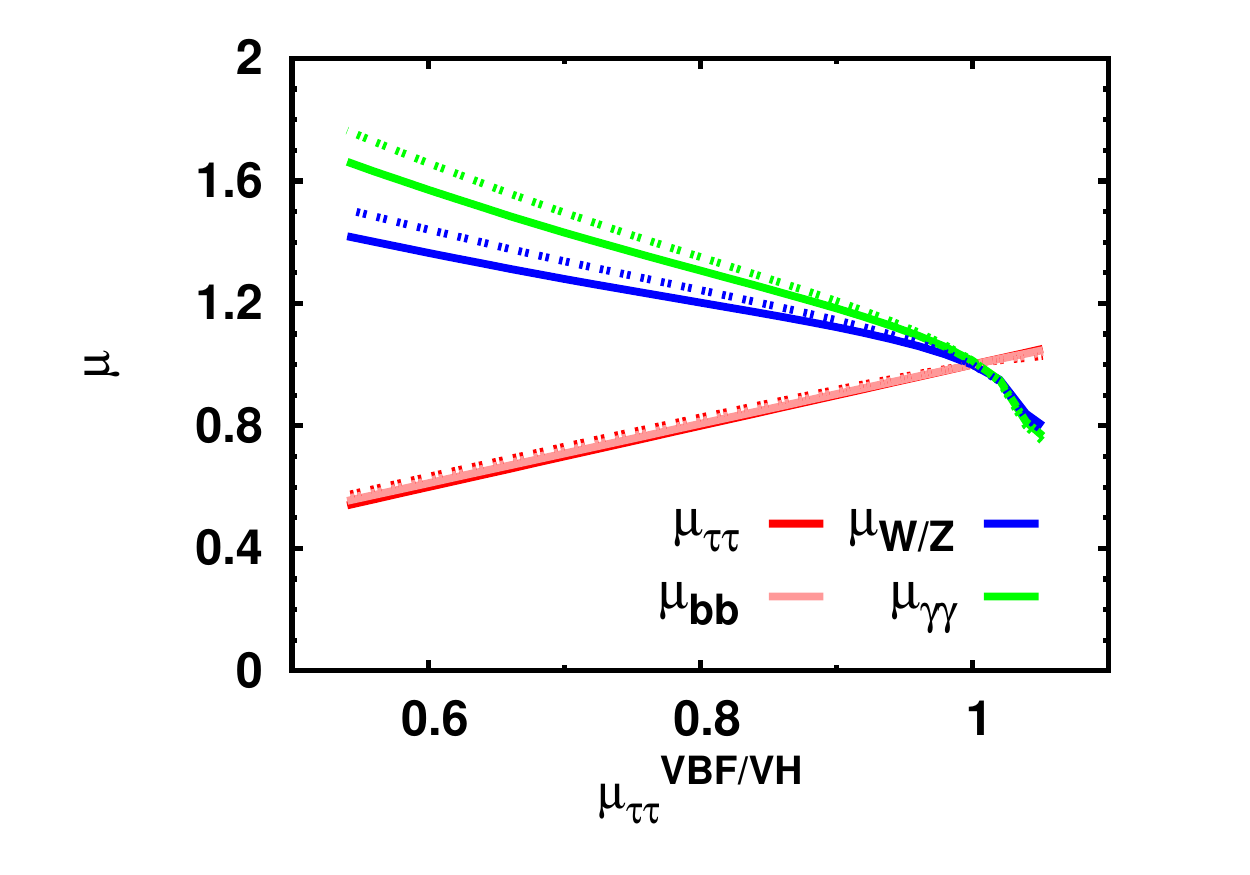} &         
\includegraphics[width=0.45\textwidth]{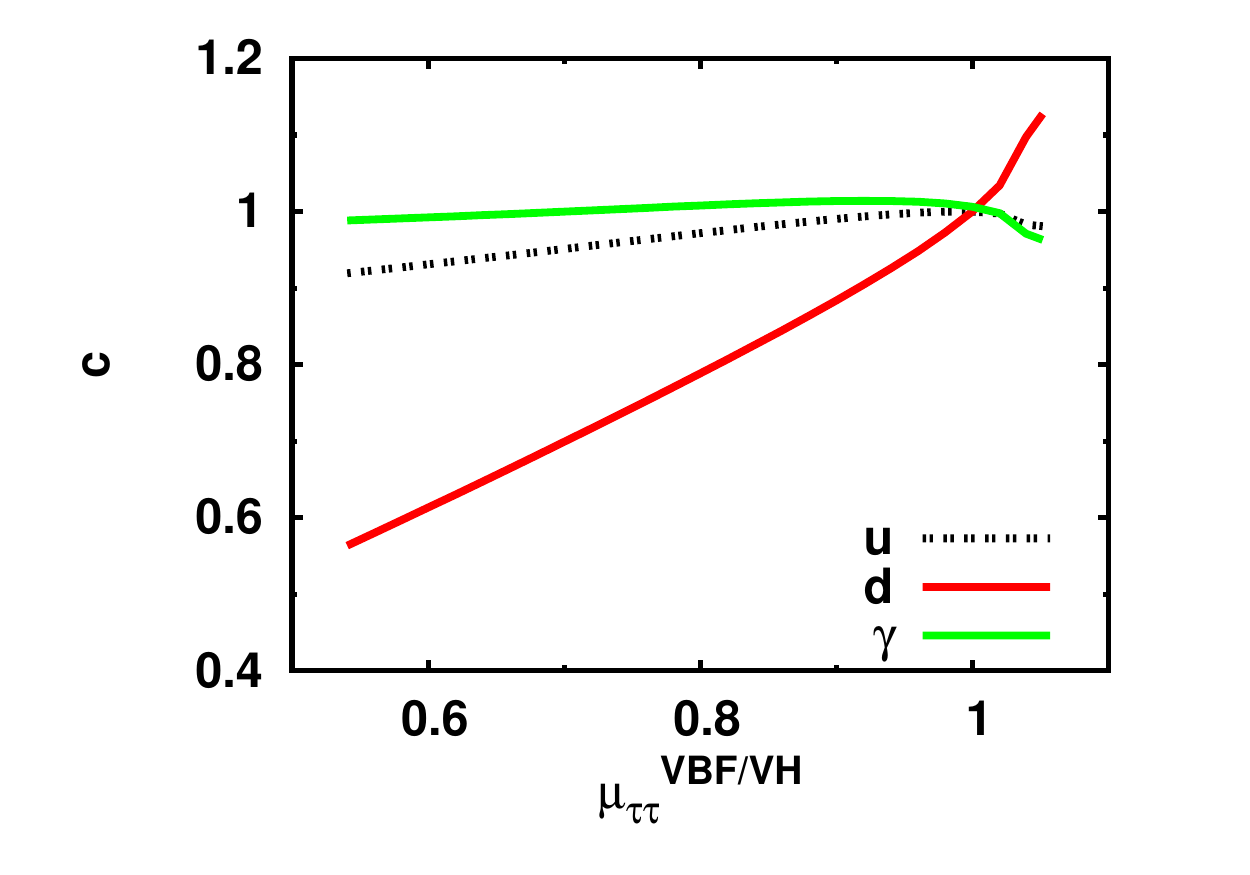} \\
(a) & (b) \\
 \includegraphics[width=0.45\textwidth]{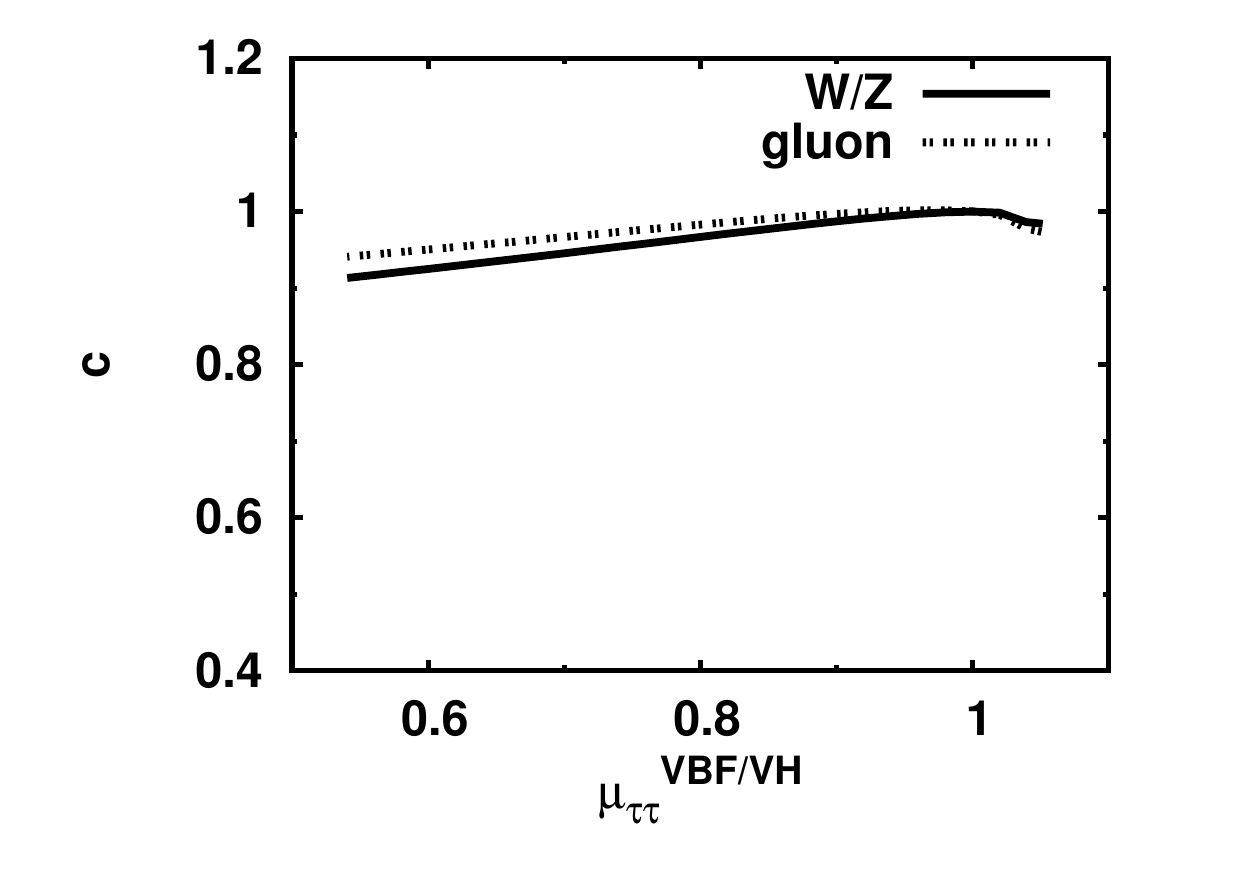} &         
 \includegraphics[width=0.45\textwidth]{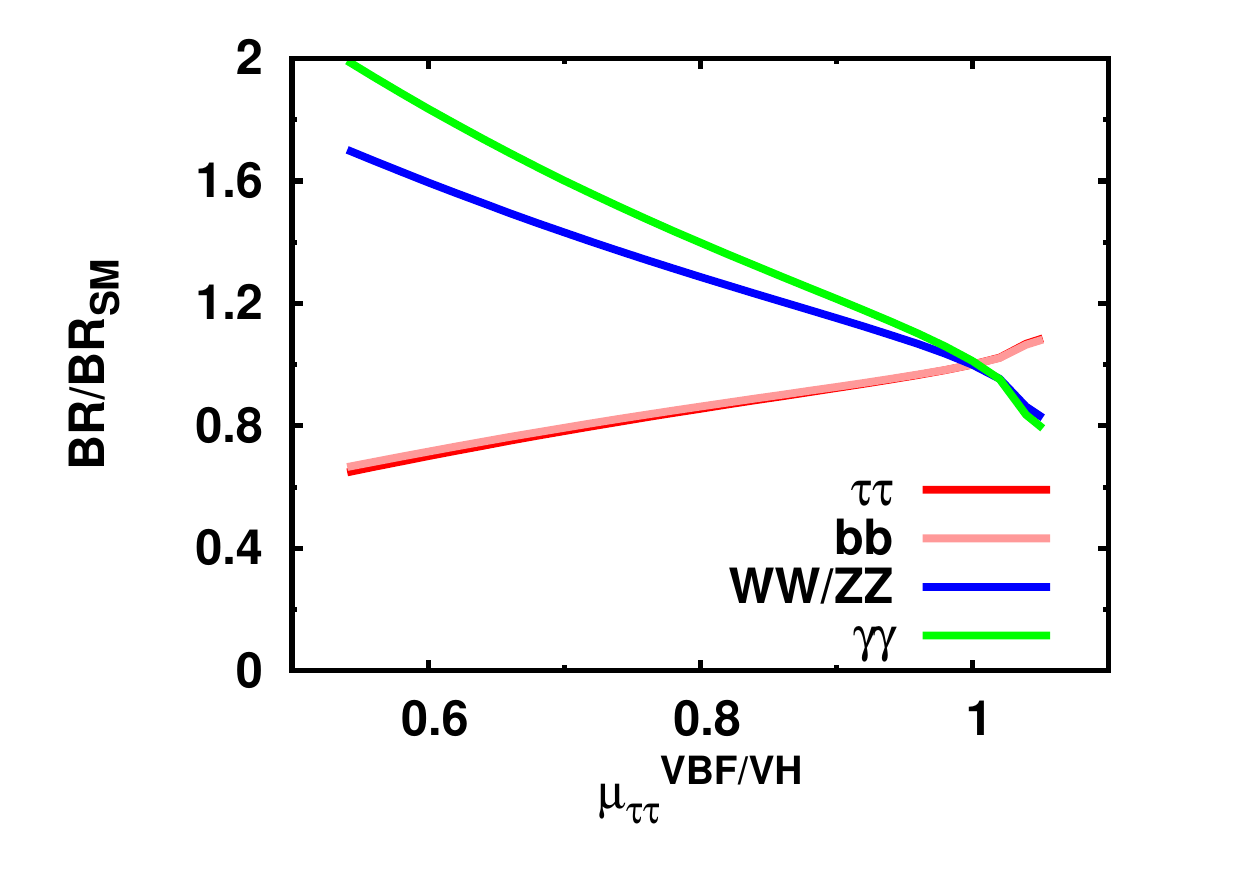} \\
(c) & (d) \\
\end{tabular}
\end{minipage}
\caption[]{ As Fig. \ref{fig:5}, but for CASE II.  In this case the change in signal strengths in panel (a) originates from the change in reduced couplings,  shown in panels (b) and (c). The variation of the signal strengths is obtained in the fit by varying predominatly the coupling to down-type fermions $c_d$ (lowest line in panel (b)), which varies predominantly the corresponding BRs (lowest line in panel (d)). The decrease of the BR to down-type fermions can be compensated by an increase in the \textit{bosonic} BRs, since the sum of all BRs stays almost constant, if the total width stays almost constant. The anti-correlation in BRs leads to anti-correlations in the signal strengths, as shown in panel (a) by the lines with negative (positive) slopes for \textit{bosonic} (\textit{fermionic}) signal strengths. 
}
\label{fig:6}
\end{center}
\end{figure} 
\begin{figure}[]
\begin{center}
\begin{minipage}{\textwidth}
\begin{tabular}{cc}
\includegraphics[width=0.45\textwidth]{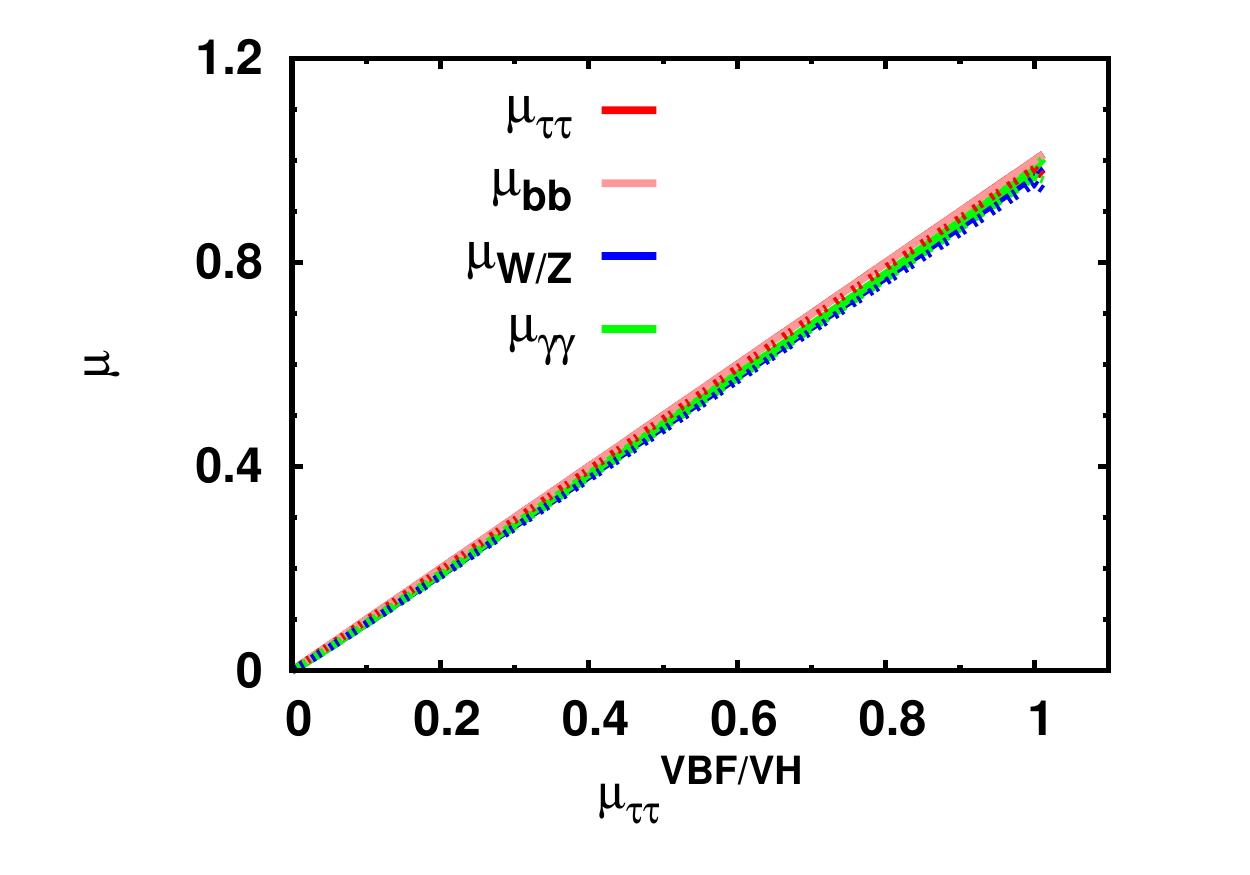} &        
\includegraphics[width=0.45\textwidth]{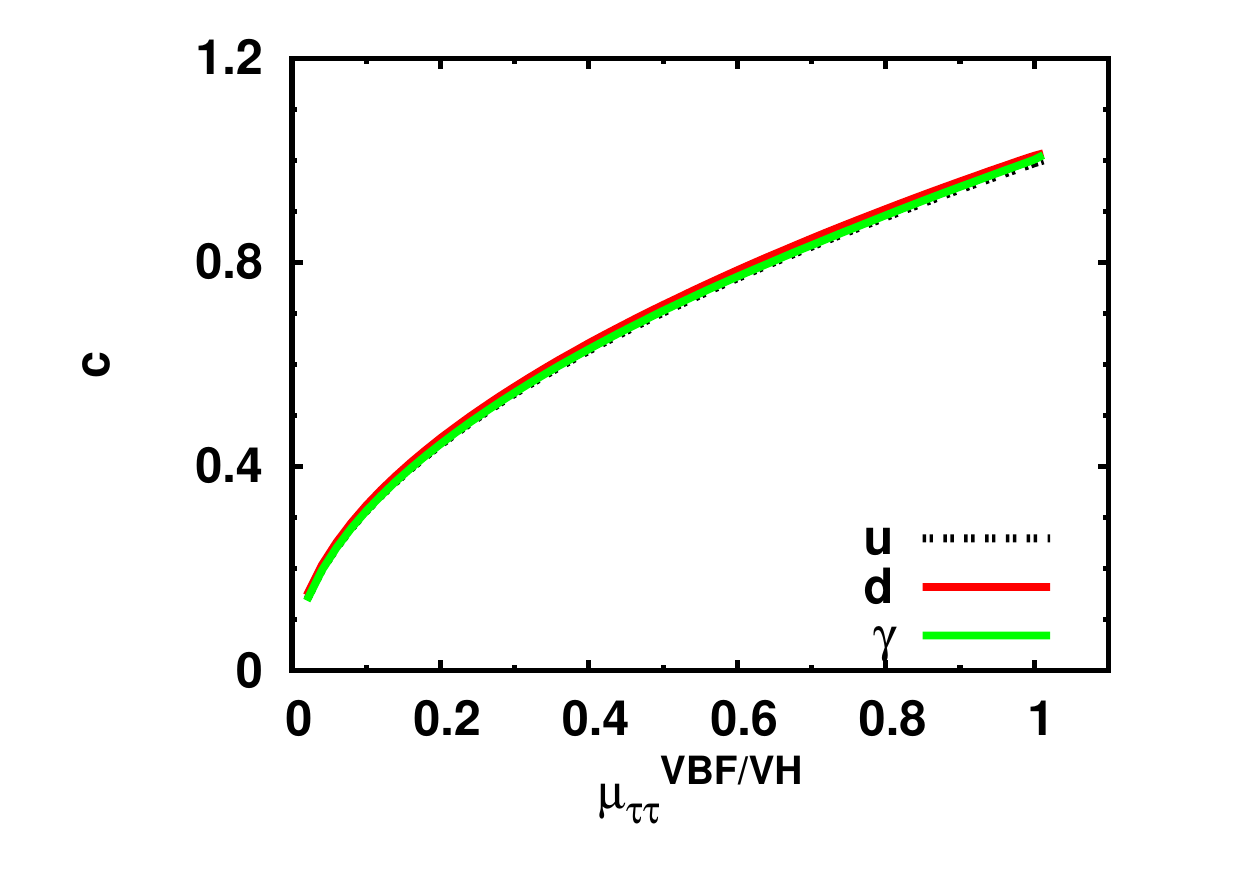}\\
(a) & (b) \\
       \includegraphics[width=0.45\textwidth]{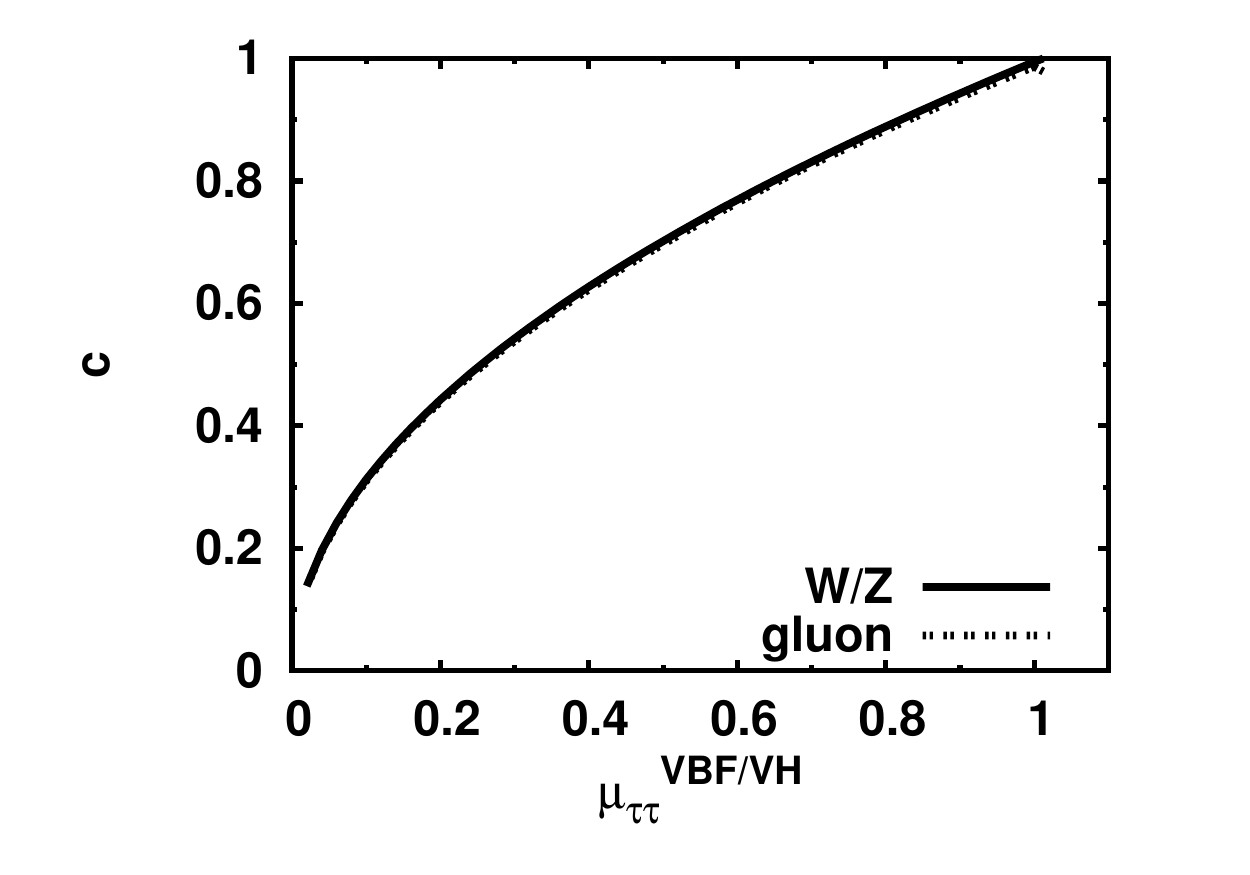} &            
        \includegraphics[width=0.45\textwidth]{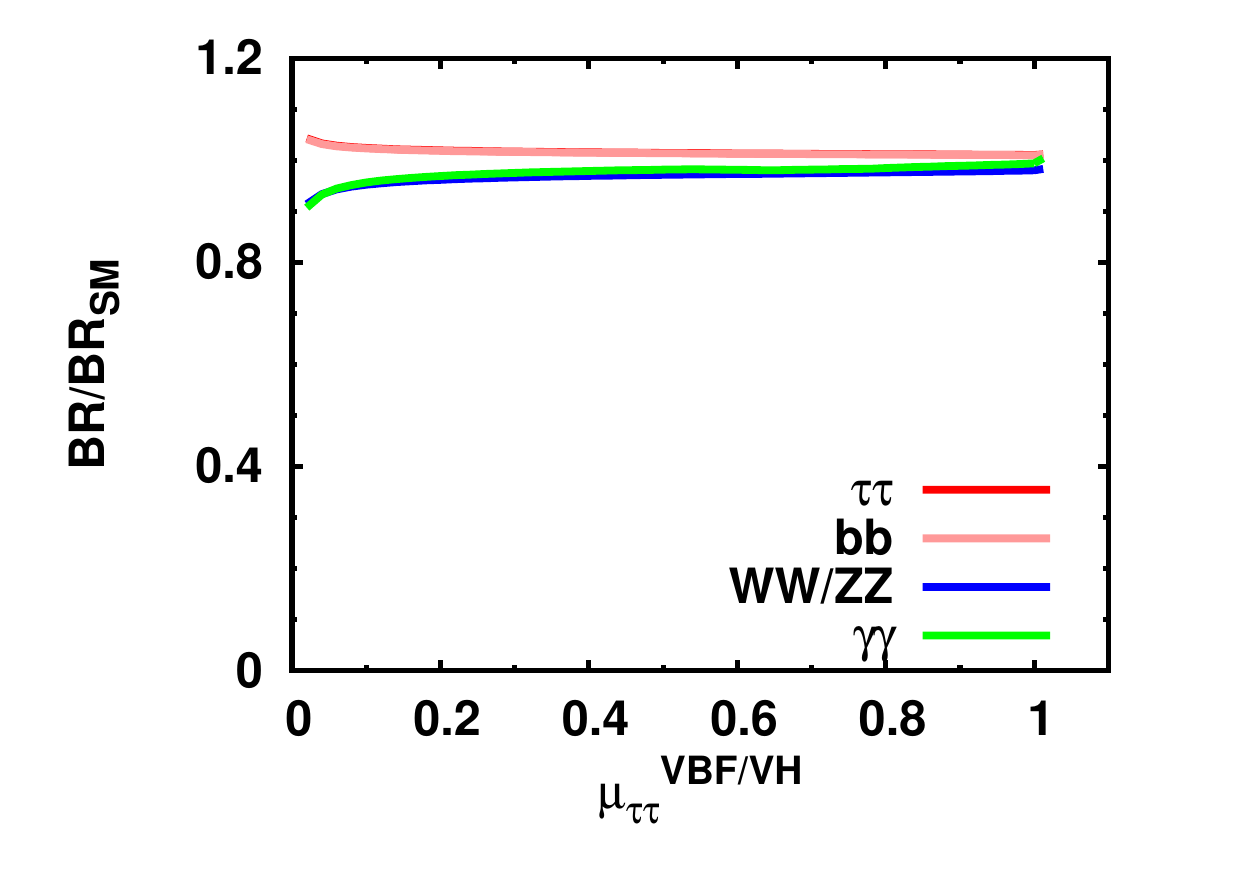}\\
(c) & (d) \\
\end{tabular}
\end{minipage}
\caption[]{
As Fig. \ref{fig:5}, but for CASE III. The variation of the signal strengths originates from a variation of  all reduced couplings simultaneously (see (b) and (c)), while the BRs stay close to 1 (see (d)). The simultaneous variation of all reduced couplings is caused by the increase of the singlet component $S_{2s}$ of the 125 GeV Higgs boson. In this case all signal strengths change in a correlated way (see overlapping lines in (a)). 
}
\label{fig:7}
\end{center}
\end{figure}  

\section{Examples of signal strength deviations of the 125 GeV Higgs bosons from the SM expectations}
\label{single}

\subsection{CASE I: Deviations by decays of the 125 GeV Higgs boson into non-SM light particles}

For CASE I either $m_{\tilde{\chi}_1^0}$ and/or $m_{A_1}$ and/or $m_{H_1}$ had to be smaller than about 60 GeV to allow decays of the 125 GeV Higgs boson into pairs of these light particles, which leads to deviations from the SM expectation. Here we  investigate the deviations from the SM for a specific mass combination in the grid of Fig. \ref{fig:2} characterized by allowed decays into neutralinos (CASE Ia), in this case $m_{H1}=90~\mathrm{GeV},m_{H3}=2000~\mathrm{GeV}$ and $m_{A1}=200~\mathrm{GeV}$. As before, the signal strength $\mu_{\tau\tau}^{VBF/VH}$ is  required to deviate from the SM expectation by  fitting it to a value $\mu_{theo} \neq 1$. This is accomplished in the fit by increasing the invisible BR via the decrease of the neutralino mass, which  can be changed in the fit to a specific value by adjusting the free parameters $\mu_{eff}$, $\lambda$ and $\kappa$ ($m_{\tilde{\chi}_0^1}\sim 2\frac{\kappa}{\lambda}\mu_{eff}$). This can be observed from the NMSSMTools output in  Appendix \ref{output} in Tables \ref{t1}-\ref{t6} for two examples, called P1 and P2, for $\mu_{theo}=1$ and 0.7, respectively.
For CASE Ib ($m_{A_1}<60$ GeV) and CASE Ic ($m_{H_1}<60$ GeV) one cannot study the deviations for a fixed mass combination, since they require changes in $m_{A_1}$ and/or $m_{H_1}$, which contradicts the study for a fixed mass combination. 

In the top left panel of Fig. \ref{fig:5} the signal strengths are shown for all 8 signal strengths as function of $\mu_{sel}=\mu_{\tau\tau}^{VBF/VH}$ between 0.3 and 1 by the overlapping solid (dashed) lines for the VBF/VH (ggf) production mode. Note that for the $b\bar{b}$ final states the ttH production mode   is selected instead of  ggf, as indicated in Eq. \ref{coupling5} before. Below $\mu_{theo}=0.3$ no good fit can be obtained. All signal strengths vary in a correlated way, as expected for CASE I. Since the signal strengths are the product of couplings squared and BRs relative to the SM values, we check which one is varying. 
Fig. \ref{fig:5}(b) shows the reduced coupling $c$ as function of the selected signal strength $\mu_{\tau\tau}^{VBF/VH}$ for $c_u$, $c_d$ and $c_\gamma$, which are all close to 1. The same is true for $c_{W/Z}$ and $c_{gluon}$ in Fig. \ref{fig:5}(c). However, from Fig. \ref{fig:5}(d) one observes that the correlated change in the signal strengths in Fig. \ref{fig:5}(a) originates from the correlated change of  BRs into visible final states (overlapping lines with a positive slope), which are anti-correlated with the invisible BR (line with negative slope). 
For $\mu_{\tau\tau}^{VBF/VH}=1$ the invisible BR is zero, while all other reduced BRs  are equal 1, so no deviations from the SM expectations are observed.
Note that experimentally the signal strengths are constrained by the data, but they are not yet precise, as shown in Fig. \ref{signal} of  Appendix \ref{error} for the data from the ATLAS and CMS experiments, which were combined by the Particle Data Group.\cite{Tanabashi:2018oca}

\subsection{CASE II: Deviations by  Higgs mixing}

For CASE II  a mass combination with all particles above 60 GeV is selected, so no decays into light particles can occur, in this case the mass combination $m_{H1}=90~\mathrm{GeV}, m_{H3}=1000~\mathrm{GeV}$ and $m_{A1}=200~\mathrm{GeV}$ was selected.  As before, regions with deviations from the SM expectations are searched for by looking for a good $\chi^2$ value under the constraint $\mu_{sel}=\mu_{\tau\tau}^{VBF/VH} =\mu_{theo}<1$.  The fit accomplishes this by modifying the NMSSM parameters leading to deviations in the Higgs mixing,  as shown in  Appendix \ref{output} in Table \ref{t1} for two selected mass combinations, P3 and P4, for $\mu_{theo}=1$ and 0.7, respectively. The change in mixing can be observed e.g. by the change in  $S_{2s}$ in Table \ref{t3} for P4 in comparison with P3. 

By fitting the down-type signal strength $\mu_{\tau\tau}^{VBF/VH}$ to $\mu_{theo}< 1$, the reduced coupling $c_d$ decreases, as shown in Fig. \ref{fig:6}(b) (lowest line), while all other reduced couplings, shown in Fig. \ref{fig:6}(b) and \ref{fig:6}(c), vary less by the variation of the mixing. The reduction in $c_d$ with much flatter dependencies for $c_u$ and $c_{W/Z}$ can be easily derived from Eq. \ref{coupling3} for larger values of $\tan\beta$. For the reduced couplings $c_{gluon}$ and $c_\gamma$ additional deviations from 1 can be caused by SUSY contributions in the loops. These are small in this case, where the lightest stop mass is about 1.2 TeV, as shown in Table \ref{t2} of  Appendix \ref{output}.  
The decrease in $c_d$ leads to decreasing BRs into down-type fermion final states, which are displayed in Fig. \ref{fig:6}(d) as function of $\mu_{\tau\tau}^{VBF/VH}$ (overlapping lines with a positive slope). The total width stays almost constant, if no new decay modes open up, so the sum of the BRs stays almost constant, implying a decrease of the \textit{fermionic} BRs must be  compensated by a larger BR for \textit{bosonic} final states (lines with negative slopes in panel (d)), leading to an anti-correlation of the corresponding signal strengths, as demonstrated in Fig. \ref{fig:6}(a): The \textit{fermionic} signal strengths $\mu_{\tau\tau},\mu_{bb}$ (lines with  positive slopes) follow $\mu_{sel}$, while the \textit{bosonic} signal strengths $\mu_{W/Z},\mu_{\gamma\gamma}$ (lines with  negative slopes) increase with decreasing $\mu_{\tau\tau}^{VBF/VH}$.
The anti-correlations  in signal strengths and BRs is demonstrated numerically  by comparing e.g. the signal strengths (BRs) for $H_2 \rightarrow bb$ and $H_2 \rightarrow ZZ/WW$ in Tables \ref{t5} and \ref{t6} of Appendix \ref{output}  for P4. 
This anti-correlation is in contrast to CASE I, shown in Fig. \ref{fig:5}, where all signal strengths decrease and follow approximately $\mu_{\tau\tau}^{VBF/VH}$. In addition, the reduced couplings, especially $c_d$, vary significantly, while for CASE I the reduced couplings equal 1 as function of $\mu_{\tau\tau}^{VBF/VH}$.

\subsection{CASE III: Deviations by   strong mixing with the singlet}

In CASE III we select $m_{H1}=122~\mathrm{GeV},m_{H3}=1300~\mathrm{GeV},m_{A1}=200~\mathrm{GeV}$, so $H_1$ is close in mass to $H_2$ (125 GeV), which can lead to   a stronger mixing between $H_1$ and $H_2$ than in  CASE II.  As for the previous cases, we force $\mu_{\tau\tau}^{VBF/VH}$ to deviate from 1 by requiring $\mu_{sel}=\mu_{\tau\tau}^{VBF/VH}=\mu_{theo}<1$. The required deviation in the fit is accomplished by increasing the mixing between $H_1$ and $H_2$, as demonstrated in Table \ref{t3} of Appendix \ref{output} for the selected mass combinations, P5 and P6, for $\mu_{theo}=1$ and 0.7, respectively. One observes that the singlet component   $H_{2s}$ of the 125 GeV Higgs boson becomes 0.56 for P6, while it is 0.008 for P5.
 The singlet component  does not couple to SM particles, so the  couplings to all \textit{fermions} and \textit{bosons} decrease from 1 for P5  to 0.83  for   P6 with $\mu_{\tau\tau}^{VBF/VH}=0.7$,  as shown in Table \ref{t4} of Appendix \ref{output} in the $H_2$ block. The dependencies of the reduced couplings on $\mu_{\tau\tau}^{VBF/VH}$ are displayed in  Figs. \ref{fig:7}(b) and   \ref{fig:7}(c).  In contrast to CASE I (Fig. \ref{fig:5}) and CASE II (Fig. \ref{fig:6}) the BRs of the 125 GeV Higgs boson stay close to the SM expectation of 1 as function of $\mu_{\tau\tau}^{VBF/VH}$, as shown in Fig. \ref{fig:7}(d). The constant BRs and decreasing couplings lead to correlated deviations for the signal strengths, as shown in Fig. \ref{fig:7}(a) for \textit{fermionic} and \textit{bosonic} final states.

\section{Conclusion}
Examples for signal strengths deviating from the SM expectation are presented and the correlations between the signal strengths for \textit{fermionic} and \textit{bosonic} final states have been determined. We find three different cases to obtain signal strengths deviating from the SM prediction, i.e. deviating from 1.
For the first case, additional decays of the 125 GeV Higgs boson with SM couplings into particles with $m < 0.5 m_{Higgs}$ are possible (CASE I). This leads to a modification of the total width, which changes all BRs in a correlated way, and hence leads to correlated deviations of the signal strengths, as summarized in Fig. \ref{fig:5}. In the second case, a modification of the Higgs mixing changes preferentially the reduced couplings to down-type fermions and thus the corresponding BRs (CASE II). The total width stays almost constant  by a modification of the Higgs mixing, so the sum of the BRs stays constant. Hence the decrease of  BRs into \textit{fermionic} final states has to be compensated by an increase of  BRs into \textit{bosonic} final states. This leads to anti-correlated deviations of the \textit{fermionic} and \textit{bosonic} signal strengths for CASE II, as summarized in Fig. \ref{fig:6}.
In the third case the singlet-like and SM-like Higgs bosons are rather close in mass, which allows for a strong mixing. In this case the fit leads to  a significant enhancement of the singlet component of the 125 GeV Higgs boson (CASE III), which  reduces the couplings to SM particles for all final states and thus leads to correlated deviations of \textit{fermionic} and \textit{bosonic} signal strengths, as summarized in Fig. \ref{fig:7}. 
The three different cases can be related to the corresponding regions in the Higgs mass space, as shown by the projections onto the $m_{A_1}$ and $m_{H_S}$ axes in Figs. \ref{fig:3} and \ref{fig:4}.  These projections are largely independent of the third mass $m_{H_3}$ in the Higgs mass space in Fig. \ref{fig:2}.

Allowed upper limits on the possible deviations of the signal strengths are proportional to  the measured  upper limits for the deviations from the SM predictions, which are  presently still large. Precision measurements at a linear collider would allow to search for deviations of the 125 GeV Higgs boson signal strengths with much higher precision. This would allow to study correlations of possible deviations in much more detail and either strongly constrain the NMSSM parameter space or point to preferred regions of mass space in case correlated or anti-correlated deviations between \textit{fermionic} and \textit{bosonic} signal strengths from SM expectations are found.


\providecommand{\href}[2]{#2}\begingroup\raggedright\endgroup

\input{appendix}

\end{document}

%% file: config.tex
\newlength{\dslashwidth}

\newcommand{\beq}{\begin{equation}} 
\newcommand{\eeq}{\end{equation}}
\newcommand{\beqa}{\begin{eqnarray}} 
\newcommand{\eeqa}{\end{eqnarray}}
\newcommand{\newc}{\newcommand}
\newcommand{\bq}{\begin{equation}}
\newcommand{\eq}{\end{equation}}
\newcommand{\ba}{\begin{array}}
\newcommand{\ea}{\end{array}}
\newcommand{\bqa}{\begin{eqnarray}}
\newcommand{\eqa}{\end{eqnarray}}

\newcommand{\lnf}{{\ifmmode \Lambda^{(N_f)} \else $\Lambda^{(N_f)}$\fi}}
\newcommand{\ms}{{\ifmmode \overline{MS} \else $\overline{MS}$\fi}}
\newcommand{\dr}{{\ifmmode \overline{DR} \else $\overline{DR}$\fi}}
\newcommand{\lms}{{\ifmmode \Lambda^{(5)}_{\overline{MS}} \else $\Lambda^{(5)}_{\overline{MS}}$\fi}}
\newcommand{\lam}{{\ifmmode \Lambda \else $\Lambda$\fi}}
\newcommand{\mev}{{\ifmmode {\rm MeV} \else ${\rm MeV}$\fi}}
\newcommand{\gev}{{\ifmmode {\rm GeV} \else ${\rm GeV}$\fi}}
\newcommand{\gevc}{{\ifmmode {\rm GeV/c^2} \else ${\rm GeV/c^2}$\fi}}
\newcommand{\tev}{{\ifmmode {\rm TeV} \else ${\rm TeV}$\fi}}
\newcommand{\tevc}{{\ifmmode {\rm TeV/c^2} \else ${\rm TeV/c^2}$\fi}}
\newcommand{\lp}{{\ifmmode L^+  \else $L^+$\fi}}
\newcommand{\lm}{{\ifmmode L^-  \else $L^-$\fi}}
\newcommand{\mlp}{{\ifmmode M(L^-) \else $M(L^-)$\fi}}
\newcommand{\mlz}{{\ifmmode M(L^0) \else $M(L^0)$\fi}}
\newcommand{\lz}{{\ifmmode L^0 \else $L^0$\fi}}
\newcommand{\ev}{{\ifmmode GeV/c^2 \else $GeV/c^2$\fi}}
\newcommand{\tri}{{\ifmmode \triangleup \else $\triangleup$\fi}}
\newcommand{\unl}{{\ifmmode U_{lL^0} \else $U_{lL^0}$\fi}}\newcommand{\gL}{{\ifmmode g_L \else $g_{L}$\fi}}
\newcommand{\gR}{{\ifmmode g_R  \else $g_{R}$\fi}}
\newcommand{\gumu}{{\ifmmode \gamma^{\mu} \else $\gamma^{\mu}$\fi}}
\newcommand{\gunu}{{\ifmmode \gamma^{\nu} \else $\gamma^{\nu}$\fi}}
\newcommand{\gdmu}{{\ifmmode \gamma_{\mu} \else $\gamma_{\mu}$\fi}}
\newcommand{\gdnu}{{\ifmmode \gamma_{\nu} \else $\gamma_{\nu}$\fi}}
\newcommand{\stw}{{\ifmmode\sin^2\theta_W \else $\sin^{2}\theta_{W}$ \fi}}
\newcommand{\sws}{{\ifmmode \;\sin^2\theta_W  \else $\;\sin^{2}\theta_{W}$ \fi}}
\newcommand{\cws}{{\ifmmode \;\cos^2\theta_W  \else $\;\cos^{2}\theta_{W}$ \fi}}
\newcommand{\sw}{{\ifmmode \;\sin\theta_W  \else $\sin\theta_{W}$ \fi}}
\newcommand{\cw}{{\ifmmode \;\cos\theta_W  \else $\;\cos\theta_{W}$ \fi}}
\newcommand{\tw}{{\ifmmode \;\tan\theta_W  \else $\;\tan\theta_{W}$ \fi}}
\newcommand{\qq}{{\ifmmode q\overline{q} \else $q\overline{q}$\fi}}
\newcommand{\lR}{{\ifmmode l_R \else $l_R$\fi}}
\newcommand{\lL}{{\ifmmode l_L \else $l_L$\fi}}
\newcommand{\nt}{{\ifmmode \nu_{\tau} \else $\nu_{\tau}$\fi}}
\newcommand{\nuR}{{\ifmmode \nu_R  \else $\nu_R$\fi}}
\newcommand{\nuL}{{\ifmmode \nu_L  \else $\nu_L$\fi}}
\newcommand{\qR}{{\ifmmode g_R  \else $q_R$\fi}}
\newcommand{\qL}{{\ifmmode q_L  \else $q_L$\fi}}
\newcommand{\qRp}{{\ifmmode q_R'  \else $q_{R}$'\fi}}
\newcommand{\qLp}{{\ifmmode q_L'  \else $q_{L}$'\fi}}
\newcommand{\est}{{\ifmmode e^{\bf \ast} \else $e^{\bf \ast}$\fi}}
\newcommand{\lst}{{\ifmmode l^{\bf \ast} \else $l^{\bf \ast}$\fi}}
\newcommand{\must}{{\ifmmode \mu^{\bf \ast} \else $\mu^{\bf \ast}$\fi}}
\newcommand{\taust}{{\ifmmode \tau^{\bf \ast} \else $\tau^{\bf \ast}$ \fi}}
\newcommand{\pperp}{{\ifmmode p_t  \else $p_t$\fi}}
\newcommand{\et}{{\ifmmode E_t  \else $E_t$\fi}}
\newcommand{\xt}{{\ifmmode x_t  \else $x_t$\fi}}
\newcommand{\smumu}{{\ifmmode \sigma_{\mu\mu}  \else $\sigma_{\mu\mu}$ \fi}}
\newcommand{\eg}{{\ifmmode e\gamma  \else $e\gamma$\fi}}
\newcommand{\epem}{{\ifmmode e^+e^-  \else $e^+e^-$\fi}}
\newcommand{\lplm}{{\ifmmode L^+L^-  \else $L^+L^-$\fi}}
\newcommand{\pp}{{\ifmmode p\overline p  \else $p\overline p$\fi}}
\newcommand{\llz}{{\ifmmode L^0\overline{L}^0 \else $L^0\overline{L}^0$\fi}}
\newcommand{\epemt}{{\ifmmode e^+e^- \to  \else $e^+e^- \to$\fi}}
\newcommand{\eb}{{\ifmmode E_{beam}  \else $E_{beam}$\fi}}
\newcommand{\ip}{{\ifmmode pb^{-1}  \else $pb^{-1}$\fi}}
\newcommand{\upm}{{\ifmmode ^{\pm}  \else $^{\pm}$\fi}}
\newcommand{\de}{{\ifmmode ^{\circ}  \else $^{\circ}$ \fi}}
\newcommand{\appr}{{\ifmmode \sim \else $\sim$ \fi}}
\newcommand{\corresp}{{\ifmmode \stackrel{\wedge}{=} \else $\stackrel{\wedge}{=}$ \fi}}
\newcommand{\sqrts}{{\ifmmode \sqrt{s} \else $\sqrt{s}$\fi}}
\newcommand{\zz}{{\ifmmode Z^0  \else $Z^0$\fi}}
\newcommand{\mz}{{\ifmmode M_{Z}  \else $M_{Z}$\fi}}
\newcommand{\mzs}{{\ifmmode M_{Z}^2  \else $M_{Z}^2$\fi}}
\newcommand{\mw}{{\ifmmode M_{W}  \else $M_{W}$\fi}}
\newcommand{\mws}{{\ifmmode M_{W}^2  \else $M_{W}^2$\fi}}
\newcommand{\mh}{{\ifmmode M_{Higgs}  \else $M_{Higgs}$\fi}}
\newcommand{\msusy}{{\ifmmode M_{SUSY}  \else $M_{SUSY}$\fi}}
\newcommand{\msusys}{{\ifmmode M_{SUSY}^2  \else $M_{SUSY}^2$\fi}}
\newcommand{\su}{{\ifmmode SU(3)_C\otimes\- SU(2)_L\otimes\- U(1)_Y \else $SU(3)_C\otimes\A0SU(2)_L\otimes U(1)_Y$\fi}}
\newcommand{\suthree}{{\ifmmode SU(3)_C  \else $SU(3)_C$\fi}}
\newcommand{\sutwo}{{\ifmmode  SU(2)_L\otimes U(1)_Y \else $SU(2)_L\otimes U(1)_Y$\fi}}
\newcommand{\taup}{{\ifmmode \tau_{proton} \else $\tau_{proton}$\fi}}
\newcommand{\as}{{\ifmmode \alpha_{s}  \else $\alpha_{s}$\fi}}
\newcommand{\mgut}{{\ifmmode M_{GUT}  \else $M_{GUT}$\fi}}
\newcommand{\mguts}{{\ifmmode M_{GUT}^2  \else $M_{GUT}^2$\fi}}
\newcommand{\mzero}{{\ifmmode m_0        \else $m_0$\fi}}
\newcommand{\mhalf}{{\ifmmode m_{1/2}    \else $m_{1/2}$\fi}}
\newcommand{\sq}{{\ifmmode \tilde{q}    \else $\tilde{q}$\fi}}
\newcommand{\gl}{{\ifmmode \tilde{g}    \else $\tilde{g}$\fi}}
\newcommand{\mb}{{\ifmmode m_{b}    \else $m_{b}$\fi}}
\newcommand{\mt}{{\ifmmode m_{t}    \else $m_{t}$\fi}}
\newcommand{\mts}{{\ifmmode m_{t}^2    \else $m_{t}^2$\fi}}

\newcommand{\mtau}{{\ifmmode m_{\tau}  \else $m_{\tau}$\fi}}
\newcommand{\dpp}{{\ifmmode \delta_{pert} \else $\delta_{pert}$\fi}}
\newcommand{\dnp}{{\ifmmode\delta_{non-pert}\else$\delta_{non-pert}$\fi}}
\newcommand{\dew}{{\ifmmode \delta_{\rm EW}\else $\delta_{\rm EW}$\fi}}
\newcommand{\rt}{{\ifmmode R_{\tau}  \else $R_{\tau} $\fi}}
\newcommand{\rz}{{\ifmmode R_{Z}  \else $R_{Z} $\fi}}

\newcommand{\swb}{{\ifmmode \sin^2\theta_{\overline{MS}} \else $\sin^2\theta_{\overline{MS}}$\fi}}
\newcommand{\cwb}{{\ifmmode \cos^2\theta_{\overline{MS}} \else $\cos^2\theta_{\overline{MS}}$\fi}}

\newc\AIPCP[3] {{\em AIP Conf. Proc.} {\bf #1} (#2) #3}
\newc\AJ[3] {{\em Astrophys. J.} {\bf #1} (#2) #3}
\newc\AMS[3] {{\em Ann. Math. Statist.} {\bf #1} (#2) #3}
\newc\AP[3] {{\em Ann. Phys.} {\bf #1} (#2) #3}
\newc\APJ[3] {{\em Astropart. J.} {\bf #1} (#2) #3}
\newc\APP[3] {{\em Astropart. Phys.} {\bf #1} (#2) #3}
\newc\APS[3] {{\em Astrophys. J. Suppl.} {\bf #1} (#2) #3}
\newc\ARNPS[3] {{\em Ann. Rev. Nucl. Part. Sci.} {\bf C#1} (#2) #3}
\newc\BA[3] {{\em Bayesian Anal.} {\bf C#1} (#2) #3}
\newc\CPC[3] {{\em Comput. Phys. Commun.} {\bf C#1} (#2) #3}
\newc\CP[3] {{\em Contemp. Phys.} {\bf #1} (#2) #3}
\newc\EPJ[3] {{\em Euro. Phys. Journ.} {\bf C#1} (#2) #3}
\newc\JCAP[3] {{\em JCAP} {\bf #1} (#2) #3}
\newc\JHEP[3] {{\em JHEP} {\bf #1} (#2) #3}
\newc\JPG[3] {{\em J. Phys.} {\bf G #1} (#2) #3}
\newc\IJMP[3] {{\em Int. J. Mod. Phys.} {\bf A #1} (#2) #3}
\newc\MNRAS[3] {{\em Mon. Not. Roy. Astron. Soc.} {\bf #1} (#2) #3}
\newc\MPL[3] {{\em Mod. Phys. Lett.} {\bf A #1} (#2) #3}
\newc\NAR[3] {{\em New Astron. Rev.} {\bf #1} (#2) #3}
\newc\NCA[3] {{\em Nuovo Cimento} {\bf #1} (#2) #3}
\newc\NIM[3] {{\em Nucl. Instrum. Methods} {\bf #1} (#2) #3}
\newc\NIMA[3] {{\em Nucl. Instrum. Methods} {\bf A #1} (#2) #3}
\newc\NAT[3] {{\em Nature} {\bf #1} (#2) #3}
\newc\NPB[3] {{\em Nucl. Phys.} {\bf B #1} (#2) #3}
\newc\NPA[3] {{\em Nucl. Phys.} {\bf A #1} (#2) #3}
\newc\NPPS[3] {{\em Nucl. Phys. Proc. Suppl.} {\bf #1} (#2) #3}
\newc\PLB[3] {{\em Phys. Lett.} {\bf B #1} (#2) #3}
\newc\PR[3] {{\em Phys. Rep.} {\bf #1} (#2) #3}
\newc\PRL[3] {{\em Phys. Rev. Lett.} {\bf #1} (#2) #3}
\newc\PRD[3] {{\em Phys. Rev.} {\bf D #1} (#2) #3}
\newc\PRC[3] {{\em Phys. Rev.} {\bf C #1} (#2) #3}
\newc\PTP[3] {{\em Prog. Theor. Phys.} {\bf #1} (#2) #3}
\newc\RMP[3] {{\em Rev. Mod. Phys.} {\bf #1} (#2) #3 }
\newc\RPP[3] {{\em Rept. Prog. Phys.} {\bf #1} (#2) #3 }
\newc\SC[3] {{\em Science} {\bf #1} (#2) #3 }
\newc\ZPC[3] {{\em Z. Phys.} {\bf C #1} (#2) #3}
\newc\Err[3] {{\em Erratum-ibid.} {\bf #1} (#2) #3 }

%% file: appendix.tex
\newpage

\appendix

\section{Higgs and neutralino mixing matrix in the NMSSM}
\label{higgsmixing}

The neutral components from the two Higgs doublets and singlet mix to form three physical CP-even scalar ($S$) bosons and two physical CP-odd pseudo-scalar ($P$) bosons. The elements of the corresponding mass matrices at tree level read:\cite{Miller:2003ay}
\begin{eqnarray}
{\cal M}^2_{S,11}&=&M_A^2+(M_Z^2-\lambda^2v^2)\sin^22\beta,\nonumber\\
{\cal M}^2_{S,12}&=&-\frac{1}{2}(M_Z^2-\lambda^2v^2)\sin4\beta,\nonumber\\
{\cal M}^2_{S,13}&=&-\frac{1}{2}(M_A^2\sin2\beta+\frac{2\kappa\mu^2}{\lambda})\frac{\lambda v}{\mu}\cos2\beta,\nonumber\\
{\cal M}^2_{S,22}&=&M_Z^2\cos^22\beta+\lambda^2v^2\sin^22\beta,\label{mix}\\
{\cal M}^2_{S,23}&=& 2 \lambda \mu v \left[1 - (\frac{M_A \sin 2\beta}{2 \mu} )^2
-\frac{\kappa}{2 \lambda}\sin2\beta\right],\nonumber\\
{\cal M}^2_{S,33}&=& \frac{1}{4} \lambda^2 v^2 (\frac{M_A \sin 2\beta}{\mu})^2
+ \frac{\kappa\mu}{\lambda} (A_\kappa +  \frac{4\kappa\mu}{\lambda} )
 - \frac{1}{2} \lambda \kappa v^2 \sin 2 \beta,\nonumber
\label{mixings} 
\end{eqnarray}

\begin{eqnarray}
{\cal M}^2_{P,11}&=&\frac{\mu (\sqrt{2}A_\lambda+\kappa \frac{\mu}{\lambda})}{\sin 2 \beta}=M^2_A,\nonumber\\
{\cal M}^2_{P,12}&=&\frac{1}{\sqrt{2}}\left( M^2_A \sin 2 \beta - 3 \frac{\kappa}{\lambda}\mu^2 \right)\frac{v\lambda}{\mu},\\
{\cal M}^2_{P,22}&=&\frac{1}{2}\left( M^2_A \sin 2 \beta + 3 \frac{\kappa}{\lambda} \mu^2 \right ) \frac{v^2}{\mu^2}\lambda^2 \sin 2 \beta - \frac{3}{\sqrt{2} \frac{\kappa}{\lambda}\mu A_\kappa}.\nonumber
\label{mixingp}
\end{eqnarray}

One observes that the element ${\cal M}^2_{S,22}$, which corresponds to the tree-level term of the lightest MSSM Higgs boson, can be above $M^2_Z$ because of the $\lambda^2v^2\sin^2 2 \beta$ term.
The diagonal element ${\cal M}^2_{P,11}$ at tree level corresponds to the pseudo-scalar Higgs boson in the MSSM limit of small $\lambda$, so it is called $M_A$.

Within the NMSSM the singlino, the superpartner of the Higgs singlet, mixes with the gauginos and Higgsinos, leading to an additional fifth neutralino. The resulting mixing matrix reads:\cite{Ellwanger:2009dp,Staub:2010ty}

\beq\label{eq1}
{\cal M}_0 =
\left( \ba{ccccc}
M_1 & 0 & -\frac{g_1 v_d}{\sqrt{2}} & \frac{g_1 v_u}{\sqrt{2}} & 0 \\
0 & M_2 & \frac{g_2 v_d}{\sqrt{2}} & -\frac{g_2 v_u}{\sqrt{2}} & 0 \\
-\frac{g_1 v_d}{\sqrt{2}} & \frac{g_2 v_d}{\sqrt{2}} & 0 & -\mu_\mathrm{eff} & -\lambda v_u \\
\frac{g_1 v_u}{\sqrt{2}} & -\frac{g_2 v_u}{\sqrt{2}}& -\mu_\mathrm{eff}& 0 & -\lambda v_d \\
0& 0& -\lambda v_u&  -\lambda v_d & 2 \kappa s 
\ea \right)
\eeq
with the gaugino masses $M_1$, $M_2$, the gauge couplings $g_1$, $g_2$ and the Higgs mixing parameter $\mu_{eff}$ as parameters.
Furthermore, the vacuum expectation values of the two Higgs doublets $v_d$,$v_u$, the singlet $s$ and the Higgs couplings $\lambda$ and $\kappa$ enter the neutralino mass matrix. 
 
The upper left $4 \times 4$ submatrix of the neutralino mixing matrix corresponds to the MSSM neutralino mass matrix, see e.g. Ref. \cite{Martin:1997ns}. 

The neutralino mass eigenstates are obtained from the diagonalization of ${\cal M}_0$ in Eq. \ref{eq1} and are linear combinations of the gaugino and Higgsino states: 
\beq\label{eq5}
\footnotesize{
\tilde{\chi}^0_i = {\cal N}(i,1) \left | \tilde{B} \right\rangle +{\cal N}(i,2) \left | \tilde{W}^0 \right\rangle 
  +{\cal N}(i,3) \left | \tilde{H}^0_u \right\rangle +{\cal N}(i,4) \left | \tilde{H}^0_d \right\rangle+{\cal N}(i,5) \left | \tilde{S} \right\rangle.}
\eeq

Typically, the diagonal elements in Eq. \ref{eq1} dominate over the off-diagonal terms, so the neutralino masses are of the order of $M_1$, $M_2$, while  the heavier Higgsinos are of the order of the  mixing parameter $\mu_{eff}$  and ithe  the lightes (singlino-like)  neutralino is of the order of $2 \kappa s \ = \ 2 (\kappa/ \lambda) \mu_{eff}$.

\section{Experimental data on the signal strengths of the 125 GeV Higgs boson}
\label{error}

\begin{figure}
\begin{center}
\hspace{-1cm}
\includegraphics[width=0.6\textwidth]{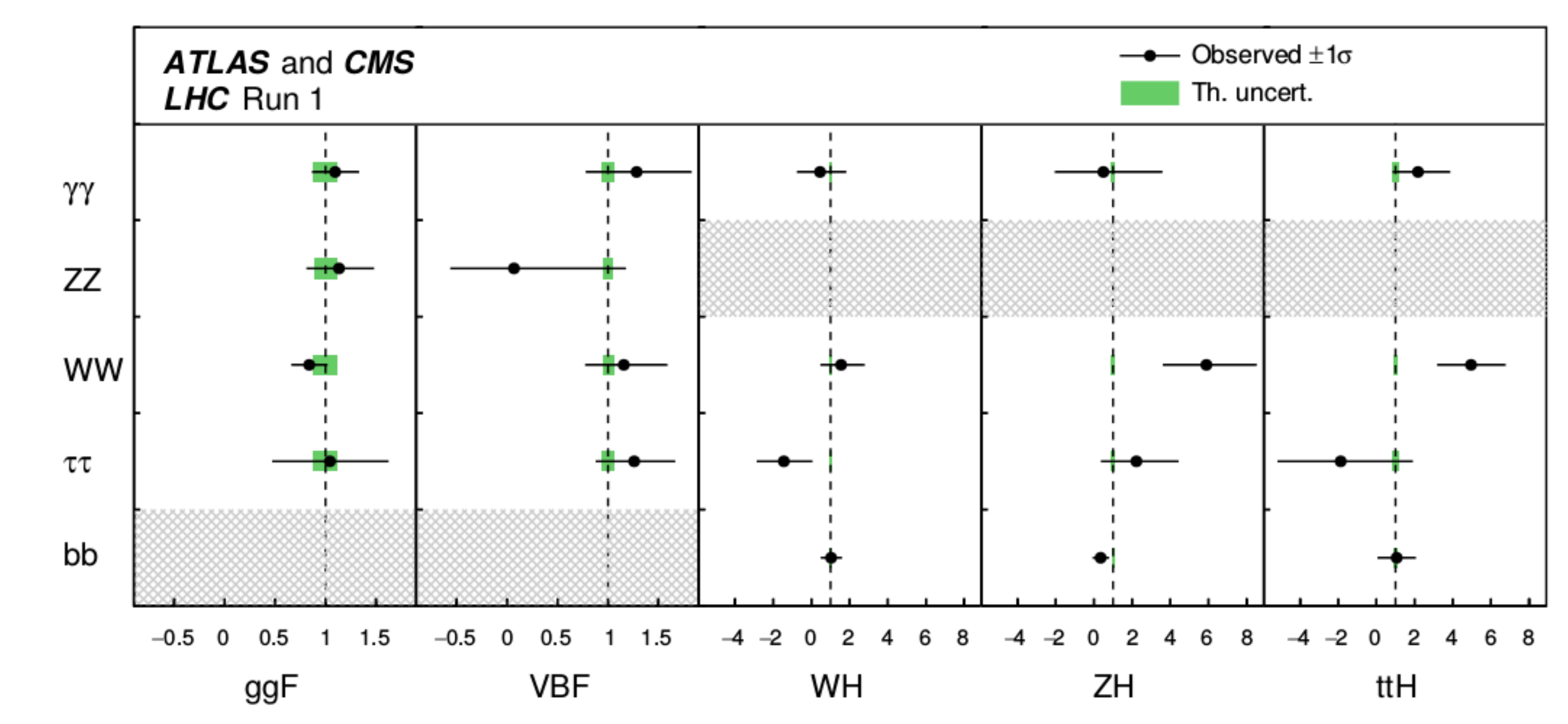}
\caption[]{ Combined measurements of the products $\sigma \cdot$ BR for the five main
production and five main decay modes. The hatched combinations require more data for a meaningful confidence interval to be provided. The figure is taken from \cite{Tanabashi:2018oca}.
}
\label{signal}
\end{center}
\end{figure}

All LHC measurements have been combined by the Particle Data Group to obtain the most precise results.\cite{Tanabashi:2018oca} A Summary of the combinations is shown in Fig. \ref{signal}.

\section{Output of NMSSMTools for  six representative  mass combinations  P1 to P6}
\label{output}
\hspace{1.5cm}

In Sect. \ref{single} examples of mass combinations for the three possible cases for deviations of  the signal strengths of the 125 GeV Higgs boson from the SM-expectation were discussed. To see which parameters need changes for deviations from the SM-expectation we present for each example the parameters without (with) deviation, i.e. $\mu_{theo}=1$($\mu_{theo}=0.7$). For CASE I we select the mass combinations $P_1$ and $P_2$,  where the last point corresponds to the deviations presented  in Fig. \ref{fig:5}. Similarly, we select $P_3$ and $P_4$ for CASE II, where $P_4$ corresponds to the deviations presented in Fig. \ref{fig:6} and  $P_5$ and $ P_6)$  for CASE III with the deviations presented in Fig. \ref{fig:7}.   
The fitted NMSSM parameters for the representative mass combinations for $H_1,H_3,A_1$ from Figs. \ref{fig:5}-\ref{fig:7} are listed in Table \ref{t1}. From these fitted parameters  NMSSMTools calculates all masses  (shown in Table \ref{t2}) and the Higgs and neutralino mixing matrices (shown in Table \ref{t3}). 
The reduced couplings are listed in Table \ref{t4}. The signal strengths and BRs are shown in \ref{t5} and \ref{t6}, respectively.  All values are obtained from the output of NMSSMTools.
     
\begin{table}
\centering
\caption{List of fitted NMSSM parameters for  representative  mass combinations  P1 to P6. \label{t1}}
\begin{tabular}{l|c|c||c|c||c|c|}
	\hline\noalign{\smallskip}
P & 1 & 2 & 3 & 4 & 5 & 6 \\
	\noalign{\smallskip}\hline\noalign{\smallskip}
$\mu_{\tau\tau}^{VBF/VH}$ & 1 & 0.7 & 1 & 0.7 & 1 & 0.7 \\
	\noalign{\smallskip}\hline\noalign{\smallskip}
	$\tan\beta$ &   4.12  & 3.96 & 5.36 & 6.57 & 12.87 &  19.47   \\ 
	$A_0$ in GeV &  -654.85 &  -553.88 & -617.09 & 520.12 & -2467.63 & -2295.48   \\
	$A_\kappa$ in GeV& 3779.52 &  3724.97 &  4673.16 &  4444.88  &  -156.54 & -156.55    \\
	$A_\lambda$ in GeV &  4325.18 & 4415.52 & 3300.37 & 3553.43 & 319.80 & -126.42  \\ 
	$\lambda \cdot 10^{-1}$&  6.31 &  6.31 & 6.97 &  6.50 & 0.04 &   0.04 \\ 
	$\kappa \cdot 10^{-1}$ &  0.37 & 0.28 & 2.51 &  3.20 & 0.03 &  0.04 \\
	$\mu_{eff}$ in GeV &   459.38 &  475.24 & 184.48 &  146.19 & 103.64 & 104.11   \\
	\noalign{\smallskip}\hline
\end{tabular}
\end{table}

\begin{table}
\centering
\caption{Masses of Higgs bosons, sparticles and gauginos in GeV for the representative  mass combinations P1 to P6. The values of   $H_1, H_3$ and $A_1$ represent the chosen mass combination in the grid of Fig. \ref{fig:2}. One observes the approximate mass degeneracy of the heavy Higgs masses ($H_3, A_2$ and  $H^\pm$). $H_2$ is the observed 125 GeV Higgs mass and $H_1$ the singlet-like boson (see large value of $S_{1s}$ in Table \ref{t3}). ). \label{t2}}
\begin{tabular}{l|c|c||c|c||c|c|}
	\hline\noalign{\smallskip}
P & 1 & 2 & 3 & 4 & 5 & 6 \\
	\noalign{\smallskip}\hline\noalign{\smallskip}
$\mu_{\tau\tau}^{VBF/VH}$ & 1 & 0.7 & 1 & 0.7 & 1 & 0.7 \\
	\noalign{\smallskip}\hline\noalign{\smallskip}
	$H_1$ & 90.0 & 90.0 & 90.0 & 90.0 &  122.9  &  122.9 \\
	$H_2$ &  125.2 & 125.2 &  125.2 & 125.2 &  125.2  &  125.3 \\ 
	$H_3$ &  2000.0 & 2000.0 & 1000.0 & 1000.0 & 1300.8   &  1300.0 \\ 
	$A_1$ & 200.0 & 200.0 & 200.0 & 200.0 & 200.0  & 200.0  \\
	$A_2$ &  2000.5 & 2000.6 & 998.1 & 997.7 & 1300.7  & 1300.0  \\
	$H^\pm$ & 1996.5 & 1996.6 &  990.1 & 990.7 &  1303.4 & 1302.7    \\
	$\tilde{d}_L$ &   2214.6 & 2214.3 & 2211.4 & 2209.9 & 2217.3 & 2218.3    \\ 
	$\tilde{d}_R$& 2137.3 & 2137.0 & 2128.1 & 2128.6 &  2134.1 &   2136.6 \\ 
	$\tilde{u}_L$ &   2213.4 & 2213.1 & 2210.2 & 2208.6 & 2216.0 &  2217.0   \\
	$\tilde{u}_R$ &  2165.3 & 2165.0 & 2184.8 & 2173.3 & 2187.0 &  2180.8    \\
	$\tilde{s}_L$ &   2214.6 & 2214.3 & 2211.4 & 2209.9 & 2217.3 & 2218.3    \\ 
	$\tilde{s}_R$&  2137.3 & 2137.0 & 2128.1 & 2128.6 & 2134.1 &  2136.6 \\ 
	$\tilde{c}_L$ &  2213.4 & 2213.1 & 2210.2 & 2208.6 & 2216.0 & 2217.0  \\
	$\tilde{c}_R$ & 2165.3 & 2165.0 &  2184.8 & 2173.3 & 2187.0 &  2180.7    \\
	$\tilde{b}_1$ &  1773.2 & 1777.3 &  1796.2 & 1884.5 & 1706.0 &  1701.9   \\ 
	$\tilde{b}_2$ & 2131.4 &  2131.6 & 2121.0 & 2122.2 &  2083.7 & 2030.4  \\ 
	$\tilde{t}_1$ & 1064.3 & 1078.1 & 1189.0 & 1427.4 &  950.8 & 1034.0   \\
	$\tilde{t}_2$ &   1792.4 & 1796.3 & 1814.5 & 1899.8 & 1731.7 & 1728.3    \\
	$\tilde{e}_L$ &   1206.3 & 1206.2 & 1233.7 & 1222.9 & 1232.0 & 1224.2   \\ 
	$\tilde{e}_R$&    1021.0 & 1021.2 & 959.2 & 986.2 & 968.0 &  987.5 \\ 
	$\tilde{\nu}^e_L$ &  1204.1 & 1204.0 & 1231.5 & 1220.6 & 1229.6 &   1221.7  \\
	$\tilde{\mu}_L$ &  1206.3 & 1206.2 &  1233.7 & 1222.9 & 1232.0 &  1224.2  \\ 
	$\tilde{\mu}_R$&  1021.0 &  1021.2 & 959.2 & 986.2 & 968.0 &  987.5  \\ 
	$\tilde{\nu}^\mu_L$ & 1204.1 & 1204.0 & 1231.5 & 1220.6 &  1229.6 & 1221.7   \\
	$\tilde{\tau}_1$ & 1016.1 & 1016.7 &  953.5 & 981.0 & 910.0  & 860.9   \\ 
	$\tilde{\tau}_2$& 1204.3 & 1204.3 &  1231.5 & 1220.8 & 1209.9  &   1176.0  \\ 
	$\tilde{\nu}^\tau_L$ &  1202.1 & 1202.1 & 1229.2 & 1218.5 & 1207.4 &  1173.3   \\
	$\tilde{g}$ &  2237.4 & 2237.3 & 2239.2 & 2240.0 & 2240.3 &  2239.3   \\
	$\tilde{\chi}^0_1 $&  62.5 &  52.1 & 103.2 & 88.8 & 98.1 &  99.0      \\
	$\tilde{\chi}^0_2 $& 407.2 & 410.7 & -220.3 &  -179.5 &  -110.9 & -111.6   \\
	$\tilde{\chi}^0_3 $&  483.9 & 494.7 & 238.2 & 224.7 & 174.8 & 174.8   \\
	$\tilde{\chi}^0_4 $&  -483.9 & -499.8 & 433.6 & 431.2 & 431.8 &  431.7  \\
	$\tilde{\chi}^0_5 $&  831.1 & 831.8 & 823.1 & 820.0 & 824.2 & 824.3   \\
	$\tilde{\chi}^\pm_1 $& 454.0 & 469.2 &  183.8 & 146.2 & 104.1 &  105.0   \\
	$\tilde{\chi}^\pm_2 $&  831.0 & 831.6 & 823.1 & 820.0 & 824.2 &  824.2  \\
	\noalign{\smallskip}\hline
\end{tabular}
\end{table}

\begin{table}
\centering
\caption{Higgs mixing matrix elements in \% for the scalar Higgs bosons $S_{ij}$ with $i=1,2,3$ and $j=d,u,s$. The Higgs mixing matrix elements for the pseudo-scalar Higgs bosons are labeled as $P_{ij}$ with $i=1,2$ and $j=d,u,s$ while the neutralino mixing matrix elements are denoted by $N_{ij}$ with $i=1,...,5$ and $j=1,...,5$. The components are defined in Eq. \ref{eq5}.
One observes that for the mass combinations P1, P3 and P5 the 125 GeV Higgs boson has a large $S_{2u}$ component and a small singlet $S_{2s}$ component. The  singlet  component $S_{2s}$  increases in CASE II and CASE III with deviations from the SM-like signal strengths (mass combinations P4 and P6 with $\mu_{\tau\tau}^{VBF/VH}$=0.7). For CASE I  the lightest neutralino (the dark matter candidate) is singlino-like with $N_{15}$=0.97, while for the other cases  with larger mixing (P3 to P6) the singlino component $N_{15}$  decreases and the Higgsino components $N_{13}$ and $N_{14}$ increase. \label{t3}}
\begin{tabular}{l|c|c||c|c||c|c|}
	\hline\noalign{\smallskip}
P & 1 & 2 & 3 & 4 & 5 & 6 \\
	\noalign{\smallskip}\hline\noalign{\smallskip}
$\mu_{\tau\tau}^{VBF/VH}$ & 1 & 0.7 & 1 & 0.7 & 1 & 0.7 \\
	\noalign{\smallskip}\hline\noalign{\smallskip}
	$S_{1d}$ &   4.64 &  4.83 &  11.41 & 15.83 & 0.12 & -2.86   \\
	$S_{1u}$ &  -1.60 & -0.59 & -2.54 & 30.50 &  0.83 & -55.51  \\
	$S_{1s}$ &  99.88 & 99.88  & 99.31 & 93.91 & 99.99 & 83.13 \\
	$S_{2d}$ & 23.62 &  24.48 & 18.42 & 10.52 &  7.86  & 4.34  \\
	$S_{2u}$ &  97.17 &  96.96 & 98.29 &  94.05 & 99.69  & 83.02 \\
	$S_{2s}$ &  0.46 & -0.62 & 0.40 & -32.32 & -0.83  &  55.58  \\
	$S_{3d}$ &   97.06 & 96.84 &  97.62 & 98.18 & 99.69  &  99.86  \\
	$S_{3u}$ & -23.57 &  -24.48 & -18.24 &  -15.00 &  -7.86 &  -5.20\\
	$S_{3s}$ &  -4.88 & -4.83 & -11.69 &  -11.68 & -0.05  &  -0.04 \\
	\noalign{\smallskip}\hline\noalign{\smallskip}
	$P_{1d}$ &   -5.08 & -5.13 & -9.43 & -9.13 & -0.05  & -0.03  \\
	$P_{1u}$ &  -1.23 & -1.29 & -1.76 & -1.39 &  $<$ -0.01  & $<$ -0.01   \\
	$P_{1s}$ &  99.86 & 99.86 &  99.54 & 99.57 &  99.99 & 99.99   \\
	$P_{2d}$ &   97.04 & 96.81 & 97.85 & 98.44 & 99.70  &  99.87  \\
	$P_{2u}$ &   23.56 &  24.46  & 18.26 & 14.98 & 7.75  & 5.13 \\
	$P_{2s}$ &    5.23 & 5.29 & 9.59 &  9.24 & 0.05  &  0.03     \\
	\noalign{\smallskip}\hline\noalign{\smallskip}
	$N_{11}$ & 2.48 &  2.30 &  8.90 & 9.48 & 9.55  &  9.34  \\
	$N_{12}$ & -2.26 & -2.12 & -7.61 & -8.27 &  -8.19 & -8.00 \\
	$N_{13}$ &   -1.83 & -2.59 & 31.09 & 41.22 & 72.51  &  72.53 \\
	$N_{14}$ &  -22.24 & -21.45 &  -62.58 & -68.91 & -67.70  & -67.73   \\
	$N_{15}$ &   97.42 & 97.59  & 70.57 & 58.26 & 0.66  & 0.68  \\
	$N_{21}$ &   83.62 & 87.64 &  -3.39 & -3.78 & -5.32 & -5.47    \\
	$N_{22}$ &   -8.55 & -7.61 &  3.98 & 4.32 & 5.78 &  5.94   \\
	$N_{23}$ & 40.14 & 35.43  & 70.42 & 70.86 &  68.76  & 68.74  \\
	$N_{24}$ &  -35.08 & -30.67 & 65.19 & 64.68 & 72.18  &  72.17     \\
	$N_{25}$ &  -9.58 & -8.03 & 27.64 & 27.60 & 0.19  &  0.19  \\
	$N_{31}$ &   -54.69 & 47.99 & 11.46 & -7.93 & -0.06 &  -0.06  \\
	$N_{32}$ &  -16.71 & 18.17 &  -7.06 & 5.09 &  0.04 &  0.05 \\
	$N_{33}$ & 57.86 &  -60.72 & 63.23 & -56.87 &  -0.61  &  -0.63   \\
	$N_{34}$ &  -57.11 & 59.69 &   -39.62 & 28.89 & 0.30  & 0.32   \\
	$N_{35}$ &   -10.95 & 10.77  & -65.20 & 76.43 & 99.99  &  99.99 \\
	$N_{41}$ &  -2.38 & -2.31 & 98.88 & 99.15 & 99.39 & 99.41 \\
	$N_{42}$ &   3.13 & 3.05  &  2.86 & 2.53 &  2.19  & 2.17 \\
	$N_{43}$ &  70.10 & 70.09 & -7.66 & -5.74 &  -3.26 &  -3.02   \\
	$N_{44}$ &  69.09 & 69.12  &  12.34 & 11.24 &  10.26 & 10.23 \\
	$N_{45}$ &   17.23 & 17.18 & 2.15 & 1.59 & $<$ 0.01  & $<$ 0.01  \\
	$N_{51}$ &  -1.89 & -1.97 & -1.22 & -1.19 & -1.10 & -1.09    \\
	$N_{52}$ &   98.15 & 97.97 & 99.34 & 99.40 & 99.47 & 99.48 \\
	$N_{53}$ & 11.07 &  11.78  &  4.27 & 3.41 & 2.05 & 1.79  \\
	$N_{54}$ &  -15.50 &  -16.07 & -10.57 & -10.32 & -9.99 & -9.98   \\
	$N_{55}$ &  -1.01 &  -1.04 & -0.40 & -0.31 &  $<$ -0.01 &  $<$ 0.01  \\
	\noalign{\smallskip}\hline
\end{tabular}
\end{table}

\begin{table}
\centering
\caption{The reduced couplings for all Higgs bosons to up-type fermions $c_u$, down-type fermions $c_d$, b-quarks $c_b$, W/Z-bosons $c_{W/Z}$, gluons $c_{gluon}$ and photons $c_\gamma$. One observes that for CASE I, represented by P1 and P2, the reduced couplings of $H_2$ are close to 1, even if the signal strength $\mu_{\tau\tau}^{VBF/VH}$=0.7.  The deviations from the SM-like signal strength in P2 is caused by the change in BRs to neutralinos, as can be seen from the last row in Table \ref{t6}. For the other cases the signal strength changes by a change of the reduced couplings,  see columns for P4 and P6 in the $H_2$ block. \label{t4}}
\begin{tabular}{ll|c|c||c|c||c|c|}
	\hline\noalign{\smallskip}
\multicolumn{2}{c|}{ P }& 1 & 2 & 3 & 4 & 5 & 6 \\
	\noalign{\smallskip}\hline\noalign{\smallskip}
\multicolumn{2}{c|}{$\mu_{\tau\tau}^{VBF/VH}$} & 1 & 0.7 & 1 & 0.7 & 1 & 0.7 \\
	\noalign{\smallskip}\hline\noalign{\smallskip}
\multirow{6}{*}{$H_1$} &	$c_u$ &  -0.017 &   -0.006 & -0.026 &  0.308 &0.008 & -0.556    \\
	& $c_d$ &   0.196 &  0.197 & 0.622 & 1.052 & 0.016 & -0.557    \\ 
	& $c_b$ &  0.196 &   0.196 &  0.622 & 1.050 & 0.016 &  -0.557  \\ 
	& $c_{W,Z}$ & -0.004 &  0.325 & -0.005 & 0.006 &   0.008 & -0.556    \\
	& $c_{gluon}$ &  0.046 & 0.038  & 0.126 & 0.291 & 0.008 & 0.553 \\
	& $c_\gamma$ &  0.053 & 0.041 &  0.139 & 0.177 & 0.007 &  0.560 \\ 
	\noalign{\smallskip}\hline\noalign{\smallskip}
\multirow{6}{*}{$H_2$} &	$c_u$ &  1.000 & 1.000 & 1.000 &  0.951 & 1.000 & 0.831  \\
	& $c_d$ & 1.001 & 0.999 & 1.004 &  0.699 &  1.014 &  0.847   \\ 
	& $c_b$ & 1.001 & 0.999 &  1.004 &  0.7000 & 1.014 &  0.847   \\ 
	& $c_{W,Z}$ &   1.000 & 1.000 & 1.000 & 0.946 & 1.000 &   0.831   \\
	& $c_{gluon}$ &  0.999 & 0.999  & 1.000 & 0.967 & 0.993 & 0.827   \\
	& $c_\gamma$ &   1.003 & 1.003  & 1.006 & 1.000 &  1.007 & 0.834   \\ 
	\noalign{\smallskip}\hline\noalign{\smallskip}
\multirow{6}{*}{$H_3$} &	$c_u$ & -0.243 & -0.252 &-0.186 & -0.152 &  -0.079 & -0.052  \\
	& $c_d$ & 4.114 & 3.953 & 5.322 &  6.524 &  12.868 &  19.471  \\ 
	& $c_b$ & 4.107 & 3.945 &  5.319 & 6.512 & 12.899 & 19.511   \\ 
	& $c_{W,Z}$ & $<$ -0.001 & $<$ -0.001 &  $<$ -0.001 & $<$ - 0.001 & -0.001 &  $<$ -0.001    \\
	& $c_{gluon}$ &  0.230 & 0.240 & 0.176 &  0.141 & 0.064 &  0.063  \\
	& $c_\gamma$ &   0.112 & 0.115 & 0.352 & 0.300 & 0.080 & 0.060   \\ 
	\noalign{\smallskip}\hline\noalign{\smallskip}
\multirow{6}{*}{$A_1$} &	$c_u$ & -0.013 & -0.013 & -0.018 & -0.014 & $<$ 0.001 & $<$ 0.001   \\
	& $c_d$ &-0.215 & -0.209 &  -0.514 & -0.607 &  -0.006  & -0.005  \\ 
	& $c_b$ &  -0.215 & -0.209  & -0.514 & -0.606 & -0.006  &  -0.005 \\ 
	& $c_{gluon}$ & 0.015  &   0.016 & 0.021 & 0.021 & $<$ 0.001  &  $<$ 0.001  \\
	& $c_\gamma$ &   0.048  & 0.046  & 0.150 & 0.195 &   0.003 &  0.003    \\ 
	\noalign{\smallskip}\hline\noalign{\smallskip}
\multirow{6}{*}{$A_2$} &	$c_u$ & 0.242  & 0.252  &  0.186 & 0.152 & 0.078 & 0.051 \\
	& $c_d$ & 4.114 &  3.952 &  5.335 & 6.542 & 12.870 & 19.472  \\ 
	& $c_b$ & 4.106 &  3.944 &  5.331 & 6.530 & 12.900  &  19.512 \\ 
	& $c_{gluon}$ & 0.272 &  0.283 & 0.226 &  0.191 &  0.128 &  0.125 \\
	& $c_\gamma$ &   0.131  & 0.135 & 0.416 & 0.353 & 0.113 & 0.100 \\ 
	\noalign{\smallskip}\hline
\end{tabular}
\end{table}

\begin{table}
\centering
\caption{Signal strengths from Eq. \ref{coupling5} for the scalar Higgs bosons $H_1, H_2$ and $H_3$ in the blocks  separated by horizontal lines.  One observes the correlated deviations of the signal strengths into \textit{fermionic} and \textit{bosonic} final states  for P2 and P6 and the anti-correlated changes in case of P4 by  comparing e.g. $H_2 \rightarrow bb$ with $H_2 \rightarrow ZZ/WW$. For P2 the deviation from the SM-like signal strength is caused by the increase into invisible states $H_2 \rightarrow invisible$ (last 2 lines of the middle block). For P4 the deviation from the SM-like signal strength is caused by a decrease of the BRs to down-type fermions, which is compensated by an increase of the BRs into bosons (compare  e.g. BR($H_2 \rightarrow bb$ and BR($H_2 \rightarrow WW$) in Table \ref{t6}).  For P6 the  deviation from the SM-like signal strength is caused by a correlated decrease of all reduced couplings, as shown in the last column ofTable \ref{t4}  in the $H_2$ block.. The explanations are given in  Sect. \ref{single}. \label{t5}}
\begin{tabular}{l|c|c||c|c||c|c|}
	\hline\noalign{\smallskip}
P & 1 & 2 & 3 & 4 & 5 & 6 \\
	\noalign{\smallskip}\hline\noalign{\smallskip}
$\mu_{\tau\tau}^{VBF/VH}$ & 1 & 0.7 & 1 & 0.7 & 1 & 0.7 \\
	\noalign{\smallskip}\hline\noalign{\smallskip}
	$VBF/VH \rightarrow H_1 \rightarrow \tau\tau$ &  $<$ 0.0001 & $<$ 0.0001 & $<$ 0.0001  & 0.1151 & $<$ 0.0001 & 0.3095 \\
	$ggf \rightarrow H_1 \rightarrow \tau\tau$ &  0.0023 & 0.0016 & 0.0173 & 0.0922 & $<$ 0.0001 &  0.3066  \\
	$VBF/VH \rightarrow H_1 \rightarrow bb$ & $<$ 0.0001 & $<$ 0.0001 &  $<$ 0.0001 & 0.1135 & $<$ 0.0001  & 0.3095  \\
	$ggf \rightarrow H_1 \rightarrow bb$ & 0.0003 & $<$ 0.0001 & 0.0007 & 0.1021 &  $<$ 0.0001  & 0.3095 \\
	$VBF/VH \rightarrow H_1 \rightarrow ZZ/WW$ & 0.0000 & 0.0000 &  0.0000 &  0.0000 & $<$ 0.0001 & 0.3081  \\
	$ggf \rightarrow H_1 \rightarrow ZZ/WW$ & 0.0000 & 0.0000 &  0.0000 &  0.0000  & $<$ 0.0001 & 0.3052 \\
	$VBF/VH \rightarrow H_1 \rightarrow \gamma\gamma$ &  $<$ 0.0001 & $<$ 0.0001 &  $<$ 0.0001 &  0.0032 & $<$ 0.0001 &  0.3124  \\
	$ggf \rightarrow H_1 \rightarrow \gamma\gamma$ &   0.0002 & $<$ 0.0001 & 0.0009 & 0.0026 & $<$ 0.0001 &  0.3095    \\
	\noalign{\smallskip}\hline\noalign{\smallskip}
	\noalign{\smallskip}\hline\noalign{\smallskip}
	$VBF/VH \rightarrow H_2 \rightarrow \tau\tau$ &  1.0009  & 0.7000 & 1.0026 & 0.7002 &  1.0100  & 0.7002     \\
	$ggf \rightarrow H_2 \rightarrow \tau\tau$ & 0.9984 &  0.6989 &  1.0026 & 0.7319 & 0.9951  &  0.6923  \\
	$VBF/VH \rightarrow H_2 \rightarrow bb$ &  1.0008  & 0.7000 & 1.0024 & 0.7100 & 1.0095  &  0.6998  \\
	$ggf \rightarrow H_2 \rightarrow bb$ & 1.0007 & 0.7001 & 1.0021 & 0.7186 & 1.0093  &   0.6997  \\
	$VBF/VH \rightarrow H_2 \rightarrow ZZ/WW$ &  0.9981 &  0.7009 & 0.9946 & 1.2802 & 0.9821 &  0.6739  \\
	$ggf \rightarrow H_2 \rightarrow ZZ/WW$ &    0.9955  & 0.6999  & 0.9946 & 1.3381 & 0.9675 & 0.6663 \\
	$VBF/VH \rightarrow H_2 \rightarrow \gamma\gamma$ & 1.0041 & 0.7057  &  1.0057 & 1.4317 & 0.9959 &  0.6788 \\
	$ggf \rightarrow H_2 \rightarrow \gamma\gamma$ &   1.0015 &  0.7047 & 1.0057 & 1.4964 & 0.9812 & 0.712  \\
	$VBF/VH \rightarrow H_2 \rightarrow invisible$ &   0.0002  &  0.2997 & 0.0000 & 0.0000 & 0.0000 & 0.0000  \\
	$ggf \rightarrow H_2 \rightarrow invisible$ &  0.0002 &   0.2993 & 0.0000 & 0.0000 & 0.0000 & 0.0000 \\
	\noalign{\smallskip}\hline\noalign{\smallskip}
	\noalign{\smallskip}\hline\noalign{\smallskip}
	$VBF/VH \rightarrow H_3 \rightarrow \tau\tau$ &$<$ 0.0001 & $<$ 0.0001 &  $<$ 0.0001 & 0.0004 &  0.0206 & 0.0149  \\
	$ggf \rightarrow H_3 \rightarrow \tau\tau$ &  69.2039 &  69.3713 & 24.7435 & 24.6346 & 68.8573 &  113.7163  \\
	$VBF/VH \rightarrow H_3 \rightarrow bb$ & $<$ 0.0001  & $<$ 0.0001 & $<$ 0.0001 & 0.0002 &  0.0142  & 0.0103 \\
	$ggf \rightarrow H_3 \rightarrow bb$ &  57.2843 & 57.0001  & 18.0317 & 18.6574 & 72.7133  & 53.4240 \\
	$VBF/VH \rightarrow H_3 \rightarrow ZZ/WW$ & $<$ 0.0001 & $<$ 0.0001  & $<$ 0.0001 & $<$ 0.0001  & $<$ 0.0001  & $<$ 0.0001 \\
	$ggf \rightarrow H_3 \rightarrow ZZ/WW$ & $<$ 0.0001  & $<$ 0.0001  &  $<$ 0.0001  & $<$ 0.0001  & $<$ 0.0001  & $<$ 0.0001    \\
	$VBF/VH \rightarrow H_3 \rightarrow \gamma\gamma$ & $<$ 0.0001  & $<$ 0.0001 & $<$ 0.0001  & $<$ 0.0001 & $<$ 0.0001   & $<$ 0.0001   \\
	$ggf \rightarrow H_3 \rightarrow \gamma\gamma$ & 0.0512 & 0.0584 & 0.1080 &  0.0520 &  0.0027  & 0.0011   \\
	$VBF/VH \rightarrow H_3 \rightarrow invisible$ & $<$ 0.0001  & $<$ 0.0001  & $<$ 0.0001  & $<$ 0.0001  & $<$ 0.0001  & $<$ 0.0001   \\
	$ggf \rightarrow H_3 \rightarrow invisible$ & 0.0013 & 0.0013 & 0.0051 & 0.0022 &  $<$ 0.0001  & $<$ 0.0001  \\
	\noalign{\smallskip}\hline
\end{tabular}
\end{table}

\begin{table}
\centering
\caption{ The BRs for the two light scalar Higgs bosons in \%. The lightest Higgs boson is predominantly singlet-like, while the second lightest Higgs boson corresponds to the 125 GeV Higgs boson, see Tables \ref{t2} and \ref{t3}.  $H_1$ decays predominantly into fermions.  Note the anti-correlation of BR($H_2 \rightarrow bb$) and 	BR($H_2 \rightarrow WW$)  for P3 and P4: the first one decreases by decreasing $\mu_{\tau\tau}^{VBF/VH}$, while the second one increases. For the other mass combinations one finds correlated changes. The explanations are given in Sect. \ref{single}. \label{t6}}
\begin{tabular}{l|c|c||c|c||c|c|}
	\hline\noalign{\smallskip}
P & 1 & 2 & 3 & 4 & 5 & 6 \\
	\noalign{\smallskip}\hline\noalign{\smallskip}
$\mu_{\tau\tau}^{VBF/VH}$ & 1 & 0.7 & 1 & 0.7 & 1 & 0.7 \\
	\noalign{\smallskip}\hline\noalign{\smallskip}
	BR($H_1 \rightarrow hadrons$) & 0.308 & 0.234 & 0.246 & 0.401 &  1.842 & 5.839   \\
	BR($H_1 \rightarrow ee$) & $<$ 0.001 & $<$ 0.001  & $<$ 0.001  & $<$ 0.001  & $<$ 0.001  & $<$ 0.001   \\
	BR($H_1 \rightarrow \mu\mu$) &  0.033 &  0.033 & 0.033 & 0.033 & 0.031  &  0.025   \\
	BR($H_1 \rightarrow \tau\tau$) & 9.315 & 9.325 &  9.301 & 9.244 & 8.872  & 7.010 \\
	BR($H_1 \rightarrow cc$) & 0.057 & 0.028 & 0.032 & 0.309 & 0.993  & 3.01 \\
	BR($H_1 \rightarrow bb$) &  90.277 & 90.375 & 90.382 & 90.007 &  81.848  &   65.278 \\
	BR($H_1 \rightarrow WW$) &  $<$ 0.001  & $<$ 0.001 & $<$ 0.001  & 0.002  & 5.729 & 16.739 \\
	BR($H_1 \rightarrow ZZ$) & -  & -  &  -  & -  & 0.588 & 1.720  \\
	BR($H_1 \rightarrow \gamma\gamma$) &  0.010 & 0.006 & 0.007 & 0.004 &  0.056 & 0.241  \\
	BR($H_1 \rightarrow Z\gamma$) & -  & -  &  - & -  & 0.041 &  0.142  \\
	\noalign{\smallskip}\hline\noalign{\smallskip}
	\noalign{\smallskip}\hline\noalign{\smallskip}
	BR($H_2 \rightarrow hadrons$) &  5.785 & 4.066 & 5.779 & 8.662 & 5.623 & 5.600 \\
	BR($H_2 \rightarrow ee$) & $<$ 0.001  & $<$ 0.001  & $<$ 0.001  & $<$ 0.001  & $<$ 0.001  & $<$ 0.001  \\
	BR($H_2 \rightarrow \mu\mu$) &  0.024 &  0.016  &  0.024 & 0.018 & 0.024  &  0.024    \\
	BR($H_2 \rightarrow \tau\tau$) & 6.651 & 4.650  & 6.660 & 5.202 &  6.708 & 6.718 \\
	BR($H_2 \rightarrow cc$) &  2.885 & 2.027  & 2.876 & 4.233 & 2.827 & 2.806 \\
	BR($H_2 \rightarrow bb$) &   61.752 & 43.177  & 61.827 & 48.973 & 62.252 & 62.334 \\
	BR($H_2 \rightarrow WW$) &  20.259 & 14.243 & 20.212 & 29.091 & 19.973 & 19.931 \\
	BR($H_2 \rightarrow ZZ$) &  2.223 & 1.563 & 2.219 & 3.193 & 2.193 & 2.192  \\
	BR($H_2 \rightarrow \gamma\gamma$) &  0.241 & 0.169 & 0.241 & 0.384 & 0.239 &  0.236   \\
	BR($H_2 \rightarrow Z\gamma$) & 0.162 & 0.114 & 0.162 & 0.244 &  0.160  & 0.159  \\
	BR($H_2 \rightarrow \tilde{\chi}^0_1\tilde{\chi}^0_1$) &    0.018 &  29.974 & - & -  & -  & - \\
	\noalign{\smallskip}\hline
\end{tabular}
\end{table}

%% file: archiv_final.bbl
\begin{thebibliography}{10}%
\makeatletter
\providecommand{\hrefCMSnoop }[0]{\@secondoftwo}%
\makeatother

\bibitem{Aad:2012tfa}
\hrefCMSnoop {} {{ ATLAS Collaboration} Collaboration, ``{Observation of a new
  particle in the search for the Standard Model Higgs boson with the ATLAS
  detector at the LHC}'',} \textit{ Phys.Lett.} \textbf{ B716} (2012) 1--29,
\href{http://www.arXiv.org/abs/1207.7214}{\texttt{ arXiv:1207.7214}}.

\bibitem{Chatrchyan:2012xdj}
\hrefCMSnoop {} {{ CMS} Collaboration, ``{Observation of a new boson at a mass
  of 125 GeV with the CMS experiment at the LHC}'',} \textit{ Phys. Lett.}
  \textbf{ B716} (2012) 30--61,
\href{http://www.arXiv.org/abs/1207.7235}{\texttt{ arXiv:1207.7235}}.

\bibitem{Carena:2012mw}
\hrefCMSnoop {} {M.~Carena, S.~Gori, I.~Low{ et~al.}, ``{Vacuum Stability and
  Higgs Diphoton Decays in the MSSM}'',} \textit{ JHEP} \textbf{ 02} (2013)
  114,
\href{http://www.arXiv.org/abs/1211.6136}{\texttt{ arXiv:1211.6136}}.

\bibitem{Ellwanger:2011aa}
\hrefCMSnoop {} {U.~Ellwanger, ``{A Higgs boson near 125 GeV with enhanced
  di-photon signal in the NMSSM}'',} \textit{ JHEP} \textbf{ 03} (2012) 044,
\href{http://www.arXiv.org/abs/1112.3548}{\texttt{ arXiv:1112.3548}}.

\bibitem{Arvanitaki:2011ck}
\hrefCMSnoop {} {A.~Arvanitaki and G.~Villadoro, ``{A Non Standard Model Higgs
  at the LHC as a Sign of Naturalness}'',} \textit{ JHEP} \textbf{ 02} (2012)
  144,
\href{http://www.arXiv.org/abs/1112.4835}{\texttt{ arXiv:1112.4835}}.

\bibitem{Gunion:2012zd}
\hrefCMSnoop {} {J.~F. Gunion, Y.~Jiang, and S.~Kraml, ``{The Constrained NMSSM
  and Higgs near 125 GeV}'',} \textit{ Phys. Lett.} \textbf{ B710} (2012)
  454--459,
\href{http://www.arXiv.org/abs/1201.0982}{\texttt{ arXiv:1201.0982}}.

\bibitem{Basso:2012tr}
\hrefCMSnoop {} {L.~Basso and F.~Staub, ``{Enhancing $h \to \gamma \gamma$ with
  staus in SUSY models with extended gauge sector}'',} \textit{ Phys. Rev.}
  \textbf{ D87} (2013), no.~1, 015011,
\href{http://www.arXiv.org/abs/1210.7946}{\texttt{ arXiv:1210.7946}}.

\bibitem{Mahmoudi:2012eh}
\hrefCMSnoop {} {F.~Mahmoudi, A.~Arbey, M.~Battaglia{ et~al.}, ``{Implications
  of LHC Higgs and SUSY searches for MSSM}'',} \textit{ PoS} \textbf{
  ICHEP2012} (2013) 124,
\href{http://www.arXiv.org/abs/1211.2794}{\texttt{ arXiv:1211.2794}}.

\bibitem{Baer:2012up}
\hrefCMSnoop {} {H.~Baer, V.~Barger, P.~Huang{ et~al.}, ``{Radiative natural
  SUSY with a 125 GeV Higgs boson}'',} \textit{ Phys. Rev. Lett.} \textbf{ 109}
  (2012) 161802,
\href{http://www.arXiv.org/abs/1207.3343}{\texttt{ arXiv:1207.3343}}.

\bibitem{Vasquez:2012hn}
\hrefCMSnoop {} {D.~A. Vasquez, G.~Belanger, C.~Boehm{ et~al.}, ``{The 125 GeV
  Higgs in the NMSSM in light of LHC results and astrophysics constraints}'',}
  \textit{ Phys.Rev.} \textbf{ D86} (2012) 035023,
\href{http://www.arXiv.org/abs/1203.3446}{\texttt{ arXiv:1203.3446}}.

\bibitem{Espinosa:2012ir}
\hrefCMSnoop {} {J.~R. Espinosa, C.~Grojean, M.~M{\"u}hlleitner{ et~al.},
  ``{Fingerprinting Higgs Suspects at the LHC}'',} \textit{ JHEP} \textbf{ 05}
  (2012) 097,
\href{http://www.arXiv.org/abs/1202.3697}{\texttt{ arXiv:1202.3697}}.

\bibitem{Choi:2019yrv}
\hrefCMSnoop {} {K.~Choi, S.~H. Im, K.~S. Jeong{ et~al.}, ``{A 96 GeV Higgs
  boson in the general NMSSM}'',}
\href{http://www.arXiv.org/abs/1906.03389}{\texttt{ arXiv:1906.03389}}.

\bibitem{Tanabashi:2018oca}
\hrefCMSnoop {} {{ Particle Data Group} Collaboration, ``{Review of Particle
  Physics}'',} \textit{ Phys. Rev.} \textbf{ D98} (2018), no.~3,
030001.

\bibitem{Djouadi:2005gi}
\hrefCMSnoop {} {A.~Djouadi, ``{The Anatomy of electro-weak symmetry breaking.
  I: The Higgs boson in the standard model}'',} \textit{ Phys. Rept.} \textbf{
  457} (2008) 1--216,
\href{http://www.arXiv.org/abs/hep-ph/0503172}{\texttt{ arXiv:hep-ph/0503172}}.

\bibitem{Djouadi:2005gj}
\hrefCMSnoop {} {A.~Djouadi, ``{The Anatomy of electro-weak symmetry breaking.
  II. The Higgs bosons in the minimal supersymmetric model}'',} \textit{ Phys.
  Rept.} \textbf{ 459} (2008) 1--241,
\href{http://www.arXiv.org/abs/hep-ph/0503173}{\texttt{ arXiv:hep-ph/0503173}}.

\bibitem{Martin:1997ns}
\hrefCMSnoop {} {S.~P. Martin, ``{A Supersymmetry primer}'',} \textit{
  Perspectives on supersymmetry II, Ed. G. Kane} (1997)
\href{http://www.arXiv.org/abs/hep-ph/9709356}{\texttt{ arXiv:hep-ph/9709356}}.

\bibitem{Carena:2015moc}
\hrefCMSnoop {} {M.~Carena, H.~E. Haber, I.~Low{ et~al.}, ``{Alignment limit of
  the NMSSM Higgs sector}'',} \textit{ Phys. Rev.} \textbf{ D93} (2016), no.~3,
  035013,
\href{http://www.arXiv.org/abs/1510.09137}{\texttt{ arXiv:1510.09137}}.

\bibitem{Ellwanger:2009dp}
\hrefCMSnoop {} {U.~Ellwanger, C.~Hugonie, and A.~M. Teixeira, ``{The
  Next-to-Minimal Supersymmetric Standard Model}'',} \textit{ Phys.Rept.}
  \textbf{ 496} (2010) 1--77,
\href{http://www.arXiv.org/abs/0910.1785}{\texttt{ arXiv:0910.1785}}.

\bibitem{Kim:1983dt}
\hrefCMSnoop {} {J.~E. Kim and H.~P. Nilles, ``{The mu Problem and the Strong
  CP Problem}'',} \textit{ Phys. Lett.} \textbf{ 138B} (1984)
150--154.

\bibitem{Miller:2003ay}
\hrefCMSnoop {} {D.~Miller, R.~Nevzorov, and P.~Zerwas, ``{The Higgs sector of
  the next-to-minimal supersymmetric standard model}'',} \textit{ Nucl.Phys.}
  \textbf{ B681} (2004) 3--30,
\href{http://www.arXiv.org/abs/hep-ph/0304049}{\texttt{ arXiv:hep-ph/0304049}}.

\bibitem{Hall:2011aa}
\hrefCMSnoop {} {L.~J. Hall, D.~Pinner, and J.~T. Ruderman, ``{A Natural SUSY
  Higgs Near 126 GeV}'',} \textit{ JHEP} \textbf{ 04} (2012) 131,
\href{http://www.arXiv.org/abs/1112.2703}{\texttt{ arXiv:1112.2703}}.

\bibitem{King:2012is}
\hrefCMSnoop {} {S.~King, M.~M{\"u}hlleitner, and R.~Nevzorov, ``{NMSSM Higgs
  Benchmarks Near 125 GeV}'',} \textit{ Nucl.Phys.} \textbf{ B860} (2012)
  207--244,
\href{http://www.arXiv.org/abs/1201.2671}{\texttt{ arXiv:1201.2671}}.

\bibitem{Kang:2012sy}
\hrefCMSnoop {} {Z.~Kang, J.~Li, and T.~Li, ``{On Naturalness of the MSSM and
  NMSSM}'',} \textit{ JHEP} \textbf{ 1211} (2012) 024,
\href{http://www.arXiv.org/abs/1201.5305}{\texttt{ arXiv:1201.5305}}.

\bibitem{Cao:2012fz}
\hrefCMSnoop {} {J.-J. Cao, Z.-X. Heng, J.~M. Yang{ et~al.}, ``{A SM-like Higgs
  near 125 GeV in low energy SUSY: a comparative study for MSSM and NMSSM}'',}
  \textit{ JHEP} \textbf{ 1203} (2012) 086,
\href{http://www.arXiv.org/abs/1202.5821}{\texttt{ arXiv:1202.5821}}.

\bibitem{Ellwanger:2012ke}
\hrefCMSnoop {} {U.~Ellwanger and C.~Hugonie, ``{Higgs bosons near 125 GeV in
  the NMSSM with constraints at the GUT scale}'',} \textit{ Adv.High Energy
  Phys.} \textbf{ 2012} (2012) 625389,
\href{http://www.arXiv.org/abs/1203.5048}{\texttt{ arXiv:1203.5048}}.

\bibitem{Beskidt:2013gia}
\hrefCMSnoop {} {C.~Beskidt, W.~de~Boer, and D.~Kazakov, ``{A comparison of the
  Higgs sectors of the CMSSM and NMSSM for a 126 GeV Higgs boson}'',} \textit{
  Phys.Lett.} \textbf{ B726} (2013) 758--766,
\href{http://www.arXiv.org/abs/1308.1333}{\texttt{ arXiv:1308.1333}}.

\bibitem{Hugonie:2007vd}
\hrefCMSnoop {} {C.~Hugonie, G.~Belanger, and A.~Pukhov, ``{Dark matter in the
  constrained NMSSM}'',} \textit{ JCAP} \textbf{ 0711} (2007) 009,
\href{http://www.arXiv.org/abs/0707.0628}{\texttt{ arXiv:0707.0628}}.

\bibitem{Kozaczuk:2013spa}
\hrefCMSnoop {} {J.~Kozaczuk and S.~Profumo, ``{Light NMSSM neutralino dark
  matter in the wake of CDMS II and a 126 GeV Higgs boson}'',} \textit{ Phys.
  Rev.} \textbf{ D89} (2014), no.~9, 095012,
\href{http://www.arXiv.org/abs/1308.5705}{\texttt{ arXiv:1308.5705}}.

\bibitem{Ellwanger:2014dfa}
\hrefCMSnoop {} {U.~Ellwanger and C.~Hugonie, ``{The semi-constrained NMSSM
  satisfying bounds from the LHC, LUX and Planck}'',} \textit{ JHEP} \textbf{
  08} (2014) 046,
\href{http://www.arXiv.org/abs/1405.6647}{\texttt{ arXiv:1405.6647}}.

\bibitem{Beskidt:2014oea}
\hrefCMSnoop {} {C.~Beskidt, W.~de~Boer, and D.~I. Kazakov, ``{The impact of a
  126 GeV Higgs on the neutralino mass}'',} \textit{ Phys. Lett.} \textbf{
  B738} (2014) 505--511,
\href{http://www.arXiv.org/abs/1402.4650}{\texttt{ arXiv:1402.4650}}.

\bibitem{Cao:2016nix}
\hrefCMSnoop {} {J.~Cao, Y.~He, L.~Shang{ et~al.}, ``{Natural NMSSM after LHC
  Run I and the Higgsino dominated dark matter scenario}'',} \textit{ JHEP}
  \textbf{ 08} (2016) 037,
\href{http://www.arXiv.org/abs/1606.04416}{\texttt{ arXiv:1606.04416}}.

\bibitem{Xiang:2016ndq}
\hrefCMSnoop {} {Q.-F. Xiang, X.-J. Bi, P.-F. Yin{ et~al.}, ``{Searching for
  Singlino-Higgsino Dark Matter in the NMSSM}'',} \textit{ Phys. Rev.} \textbf{
  D94} (2016), no.~5, 055031,
\href{http://www.arXiv.org/abs/1606.02149}{\texttt{ arXiv:1606.02149}}.

\bibitem{Beskidt:2017xsd}
\hrefCMSnoop {} {C.~Beskidt, W.~de~Boer, D.~I. Kazakov{ et~al.},
  ``{Perspectives of direct Detection of supersymmetric Dark Matter in the
  NMSSM}'',} \textit{ Phys. Lett.} \textbf{ B771} (2017) 611--618,
\href{http://www.arXiv.org/abs/1703.01255}{\texttt{ arXiv:1703.01255}}.

\bibitem{Ellwanger:2018zxt}
\hrefCMSnoop {} {U.~Ellwanger and C.~Hugonie, ``{The higgsino?singlino sector
  of the NMSSM: combined constraints from dark matter and the LHC}'',} \textit{
  Eur. Phys. J.} \textbf{ C78} (2018), no.~9, 735,
\href{http://www.arXiv.org/abs/1806.09478}{\texttt{ arXiv:1806.09478}}.

\bibitem{Beskidt:2019mos}
\hrefCMSnoop {} {C.~Beskidt and W.~de~Boer, ``{An effective scanning method of
  the NMSSM parameter space}'',}
\href{http://www.arXiv.org/abs/1905.07963}{\texttt{ arXiv:1905.07963}}.

\bibitem{Djouadi:2008yj}
\hrefCMSnoop {} {A.~Djouadi, U.~Ellwanger, and A.~M. Teixeira, ``{The
  Constrained next-to-minimal supersymmetric standard model}'',} \textit{ Phys.
  Rev. Lett.} \textbf{ 101} (2008) 101802,
\href{http://www.arXiv.org/abs/0803.0253}{\texttt{ arXiv:0803.0253}}.

\bibitem{Kowalska:2012gs}
\hrefCMSnoop {} {K.~Kowalska, S.~Munir, L.~Roszkowski{ et~al.}, ``{The
  Constrained NMSSM with a 125 GeV Higgs boson -- A global analysis}'',}
\href{http://www.arXiv.org/abs/1211.1693}{\texttt{ arXiv:1211.1693}}.

\bibitem{Das:2011dg}
\hrefCMSnoop {} {D.~Das, U.~Ellwanger, and A.~M. Teixeira, ``{NMSDECAY: A
  Fortran Code for Supersymmetric Particle Decays in the Next-to-Minimal
  Supersymmetric Standard Model}'',} \textit{ Comput.Phys.Commun.} \textbf{
  183} (2012) 774--779,
\href{http://www.arXiv.org/abs/1106.5633}{\texttt{ arXiv:1106.5633}}.

\bibitem{James:1975dr}
\hrefCMSnoop {} {F.~James and M.~Roos, ``{Minuit: A System for Function
  Minimization and Analysis of the Parameter Errors and Correlations}'',}
  \textit{ Comput.Phys.Commun.} \textbf{ 10} (1975) 343--367.

\bibitem{option}
Option MODSEL 8 2 in NMSSMTools.

\bibitem{Degrassi:2009yq}
\hrefCMSnoop {} {G.~Degrassi and P.~Slavich, ``{On the radiative corrections to
  the neutral Higgs boson masses in the NMSSM}'',} \textit{ Nucl.Phys.}
  \textbf{ B825} (2010) 119--150,
\href{http://www.arXiv.org/abs/0907.4682}{\texttt{ arXiv:0907.4682}}.

\bibitem{Staub:2010ty}
\hrefCMSnoop {} {F.~Staub, W.~Porod, and B.~Herrmann, ``{The Electroweak sector
  of the NMSSM at the one-loop level}'',} \textit{ JHEP} \textbf{ 1010} (2010)
  040,
\href{http://www.arXiv.org/abs/1007.4049}{\texttt{ arXiv:1007.4049}}.

\end{thebibliography}
